\documentclass[useAMS,usenatbib,usegraphicx,usedcolumn]{mn2e}
\usepackage{fix2col}

\newcommand{\Msun}{ M_{\odot}}
\newcommand{\Lsun}{ L_{\odot}}
\newcommand{\Rsun}{ R_{\odot}}
\newcommand{\Fsun}{ F_{\odot}}
\newcommand{\gsun}{ g_{\odot}}
\newcommand{\Myear}{M_{\odot}~{\rm yr}^{-1}}
\newcommand{\Myeararea}{M_{\odot}~{\rm yr}^{-1}R_{\odot}^{-2}}
\newcommand{\Mdot}{\dot{M}}
\newcommand{\Mdott}{\dot{M_1}}
\newcommand{\mdot}{\dot{m}}
\newcommand{\Mbh}{M}
\newcommand{\Sigmadot}{\dot{\Sigma}}

\title[Winds and the UV turnover in AGN]{Line driven winds and the UV turnover in AGN accretion discs}
\author[Ari Laor and Shane W. Davis]{Ari Laor$^{1}$\thanks{E-mail:
laor@physics.technion.ac.il (AL); swd@cita.utoronto.ca (SWD)} and Shane W. Davis$^{2}$\footnotemark[1]\\
$^{1}$Physics Department, Technion, Haifa~32000, Israel\\
$^{2}$Canadian Institute for Theoretical Astrophysics. Toronto, ON M5S3H4, Canada}
\begin{document}

\date{}

\pagerange{\pageref{firstpage}--\pageref{lastpage}} \pubyear{2002}

\maketitle

\label{firstpage}

\begin{abstract}

AGN SEDs generally show a turnover at $\lambda \sim 1000$\AA, implying
a maximal Accretion Disc (AD) temperature of $T_{\rm max}\sim 50,000$K.
Massive O stars display a similar $T_{\rm max}$, associated with
a sharp rise in a line driven mass loss $\dot{M}_{\rm wind}$ with
increasing surface temperature. AGN AD are also characterized by
similar surface gravity to massive O stars. The $\dot{M}_{\rm wind}$
of O stars reaches $\sim 10^{-5} \Myear$. Since the surface area of
AGN AD can be $10^6$ larger, the implied $\dot{M}_{\rm wind}$ in AGN AD can
reach the accretion rate $\Mdot$. A rise to $\dot{M}_{\rm
  wind}\sim\Mdot$ towards the AD center may therefore set a similar
cap of $T_{\rm max}\sim 50,000$K.  To explore this idea, we solve the
radial structure of an AD with a mass loss term, and calculate the
implied AD emission using the mass loss term derived from observations
of O stars. We find that $\Mdot_{\rm wind}$ becomes comparable to
$\Mdot$ typically at a few tens of $GM/c^2$. Thus, the standard thin AD
solution is effectively truncated well outside the innermost stable orbit.  The
calculated AD SED shows the observed turnover at $\lambda \sim
1000$\AA, which is weakly dependent on the AGN luminosity and black hole
mass. The AD SED is generally independent of the black hole spin, due
to the large truncation radius. However, a cold AD (low $\Mdot$, high
black hole mass) is predicted to be windless, and thus its SED should
be sensitive to the black hole spin.  The accreted gas may form a hot
thick disc with a low radiative efficiency inside the truncation
radius, or a strong line driven outflow, depending on its
ionization state.

\end{abstract}


\section{Introduction}
The optical-UV emission in AGN is most likely the signature of accretion on to the central massive
black hole through a thin Accretion Disc (AD, Shields 1978; Malkan 1983 and citations thereafter). 
Malkan \& Sargent (1982) noted that the UV emission 
shows a turnover characteristic of a $T_{\rm max}\sim 30,000$K blackbody. Following studies of larger samples
showed this is a general trend in AGN, where the SED shows
 a turnover from a spectral slope of $\alpha\sim -0.5$ ($F_{\nu}\propto \nu^{\alpha}$) 
at $\lambda>1000$\AA,
to  $\alpha\sim -1.5$ to $-2$ at $\lambda< 1000$\AA\ (Zheng et al. 1997; Telfer et al. 2002; 
Shang et al. 2005; Barger 
\& Cowie 2010; Shull et al. 2012; cf. Scott et al. 2004), which extends to $\sim 1$~keV (Laor et al. 1997).
The turnover at $\lambda< 1000$\AA, which corresponds to a blackbody with 
$T_{\rm max}\sim 50,000$K, 
is in contradiction with the thin local blackbody AD models, which predict
peak emission $\nu_{\rm peak}\propto (\mdot/M)^{1/4}$, where $M$ is the black hole mass, and
$\mdot\equiv L/L_{\rm Edd}$ is the luminosity in Eddington units. Thus, $\nu_{\rm peak}$ should range over more than an order of magnitude, as
broad line AGN extend over the range $\mdot=0.01-1$ and $M=10^6-10^{10}\Msun$, which is in contrast with the small
range observed (e.g. Shang et al. 2005; Davis \& Laor 2011, hereafter DL11).
For example, some models predict a peak at $\nu_{\rm peak}>10^{16}$ (e.g. Hubeny et al. 2001, DL11), while 
objects with such SEDs appear to be extremely rare (e.g. Done et al. 2012).
Furthermore, high $\mdot/M$ AD models predict significant soft X-ray thermal emission, which is also not
observed (Laor et al. 1997), which again implies the expected thermal emission from the inner hottest parts of the AD is missing. 

The extreme UV (EUV) emission spectral shape can also be constrained based on various line ratios. The analysis
of Bonning et al. (2013) of a sample of AGN reveals similar observed line ratios, again indicating similar EUV SEDs, 
and an absence of the dependence of the EUV emission on the
predicted maximum thin AD temperature in each object. 

In contrast, the SED of AD around stellar mass black holes, which peak in the
X-ray regime, matches observations remarkably well, in particular near the peak emission which
originates from the hottest innermost AD region (Davis et al. 2005; 2006). The match is accurate enough 
that it can be used to determine the black hole spin (e.g. McClintock et al. 2011).

What prevents AD in AGN from generally reaching $T_{\rm max}\gg 50,000$K?  
The universality of the observed
$\nu_{\rm peak}$ suggests it is a local process in the AGN AD atmosphere, most likely related to an atomically driven process. This process should be effective at $T\sim 50,000$~K, and absent at $T\sim 10^7$~K, relevant to AD around stellar mass black holes.

Interestingly, main sequence stars show a similar maximum temperature. The hottest O stars also do not
generally reach beyond $\sim 50,000$~K (e.g. Howarth \& Prinja 1989). Massive O stars produce a strong wind
with a high mass loss, $\Mdot_{\rm wind}$, which can reach  $10^{-5}\Myear$ in the most 
luminous O stars with a luminosity $L_*>10^6 \Lsun$. Such stars have a mass of $M_*\sim 50-100 \Msun$,
and thus loose a significant fraction of their mass on a time-scale 
$t_{\rm wind}=M_*/\Mdot_{\rm wind}\sim 5-10$~Myr,
which is comparable to their lifetime.

Is this $\Mdot_{\rm wind}$ regulation of the hottest and most massive O stars relevant to AGN? 
Can this mechanism explain the similar $T_{\rm max}$ observed in AGN AD and in O stars? 
The local structure of a stellar atmosphere is mostly set by the local flux, i.e. the effective 
temperature $T_{\rm eff}$, and by the surface gravity $g$. AGN AD have $T_{\rm eff}\sim 10^4-10^5$~K at
their inner regions,
and $g$ at the disc surface is set by the balance of radiation pressure and gravity. Thus,
the local flux and $g$ are always at the Eddington limit in the vertical direction. In O stars, the local flux and $g$ 
also reach close to the Eddington limit.
The radius of O stars is $\sim 10^{12}$~cm. The radius of the UV emitting region in luminous AGN AD is
$\sim 10^{15}$~cm, i.e. a $10^6$ larger surface area. Thus, if the mass loss per unit
surface area reaches similar values in O stars and in AGN, given their similar $T_{\rm max}$ and $g$, then the total $\Mdot_{\rm wind}$ 
from the AGN AD where $T\sim 50,000$~K can reach $10\Myear$, which can exceed the accretion rate
$\Mdot$. The value of $T_{\rm max}$ in AGN AD may then be set by the radius at which the thin disc 
solution must break down as $\Mdot_{\rm wind}>\Mdot$. 

Below we explore this suggestion more quantitatively. In \S 2 we derive the mass loss per unit
surface area in stars as a function of the atmospheric properties. In \S 3 we 
provide a simple analytic estimate
of the innermost disc radius, and the implied $T_{\rm max}$ in AGN AD, based on the
O stars mass loss. In \S 4 we derive revised equation
for the radial AD structure for a thin AD + wind. In \S 5 we provide numerical solutions
for the revised thin AD radial structure. In \S 6 we derive the AD SED using various
approximations. The results are discussed in \S 7, and the main conclusions are summarized in \S 8.

\section{Stellar mass loss}
Below we use the stellar $\Mdot_{\rm wind}(L_*)$ relation to derive a relation between
the mass loss per unit area, $\Sigmadot$, and the locally emitted flux per unit area (i.e 
$T_{\rm eff}$). We assume this relation applies to AGN AD, and use the AD expression
for $T_{\rm eff}(R)$ to derive $\Sigmadot(R)$, where $R$ is the radius. We then integrate over the AD surface area
to derive the cumulative $\Mdot_{\rm wind}(R)$, and derive $R_{\rm eq}$ at which
$\Mdot_{\rm wind}(R_{\rm eq})=\Mdot$, where $\Mdot$ is the accretion rate coming in from infinity.
This radius forms the effective inner thin AD boundary, and sets the maximum thin disc
temperature, $T_{\rm max}=T_{\rm eff}(R_{\rm eq})$.

Observations of O stars yield the following tight relation between $\Mdot_{\rm wind}$ and $L_*$,
\begin{equation}
 \log \Mdot_{\rm wind}/\Myear=1.69\log L_*/\Lsun -15.4 ,
\end{equation}
derived in the range $4.5<\log L_*/\Lsun<6.5$, (Howarth \& Prinja 1989), which corresponds to 
an effective temperature in the range $30,000<\log T_{\rm eff}<50,000$~K, where the O stars range from
main sequence to supergiants. Howarth \& Prinja (1989) list the stellar radius
$R_*$ for their sample of 201 stars, which we use to derive the mass loss rate per unit area, 
$\Sigmadot$, for each star. Figure 1 (top panel) 
presents the derive best fit linear relation of 
$\Sigmadot$ and $F/\Fsun$,
\begin{equation}
 \log \Sigmadot =1.9\log F/\Fsun-15.7 ,
\end{equation}
where $\Sigmadot$ is measured here and below in units of $\Myeararea$,  
and $F/\Fsun$ is the flux in solar flux units, which equals $(T_{\rm eff}/5774)^4$.
The relation has a scatter of 0.28 in $\log \Sigmadot$ at a given $T_{\rm eff}$ (see Figure 1).

Solutions for the atmospheric structure are set by $T_{\rm eff}$ and $g$.
Thus, although the global $\Mdot_{\rm wind}$ in stars is set by $L_*$ only, the local 
$\Sigmadot$ is likely set by both $T_{\rm eff}$ and $g$. Figure 1 (middle panel) 
shows the derived best fit relation of $\Sigmadot$
vs. $F$ and $g$ 
\begin{equation}
 \log \Sigmadot=2.32\log F/\Fsun-1.11\log g/\gsun-17.72 ,
\end{equation}
where $g/\gsun$ is the surface gravity in solar units.
Indeed, the relation is significantly tighter, and the scatter reduces to 0.06.

This relation is consistent with the line driven winds solution (Castor et al. 1975;
hereafter CAK) which yields, to a good approximation,
\begin{equation}
 \Mdot_{\rm wind} \propto [M_*(1-\Gamma)]^{(\alpha-1)/\alpha}L_*^{1/\alpha} ,
\end{equation}
where $\Gamma=L_*\sigma_{\rm es}/4\pi cGM_*$ represents $L_*$ in Eddington units, and
$\alpha$ describes the power law dependence of the force multiplier $M$ on
the electron scattering optical depth, $t$, from the surface of the atmosphere,
$M\propto t^{-\alpha}$. Since $\alpha\simeq 0.5$ (e.g. Lamers \& Cassinelli 1999), 
and $\Gamma<0.5$ for O stars (Figure 1, lower panel), we get in the limit $\Gamma\ll 1$ that 
$\Mdot_{\rm wind} \propto M_*^{-1}  L_*^2$. Now, using the local quantities, $g\propto M_*/R^2$,
and $F\propto L_*/R^2$, we derive $\Sigmadot \propto g^{-1}F^2$, which is close 
to the relation found above (equation 3). 

Although the $\Sigmadot(F,g)$ relation (equation 3) is significantly tighter than the
$\Sigmadot(F)$ relation (equation 2), its applicability to AGN AD is not clear. In the 
radiation pressure dominated part of AD, relevant to AGN AD at small $R$, 
$\Gamma=1$ at the disc surface, and the
CAK expression for $\Mdot_{\rm wind}$ (equation 4) formally diverges. However, in AD the dynamics
is different, as $g\propto z$ and $F$ is constant for $z\ll R$, in contrast with
stars where both $g$ and $F$ are $\propto 1/R^2$, so $\Gamma>1$ does not lead to divergence
as in the stellar case. We therefore use below both 
the $\Sigmadot(F,g)$ and the $g$ averaged $\Sigmadot(F)$ relation, 
to get some indication of possible $\Mdot_{\rm wind}$ values.

We note in passing that additional $\Sigmadot(F,g)$ relations can be derived
from various theoretical calculations presented by Vink et al. (2000) and Lucy (2010). For the sake
of simplicity we use only the observationally derived relations given above.

\begin{figure}
\includegraphics[width=144mm]{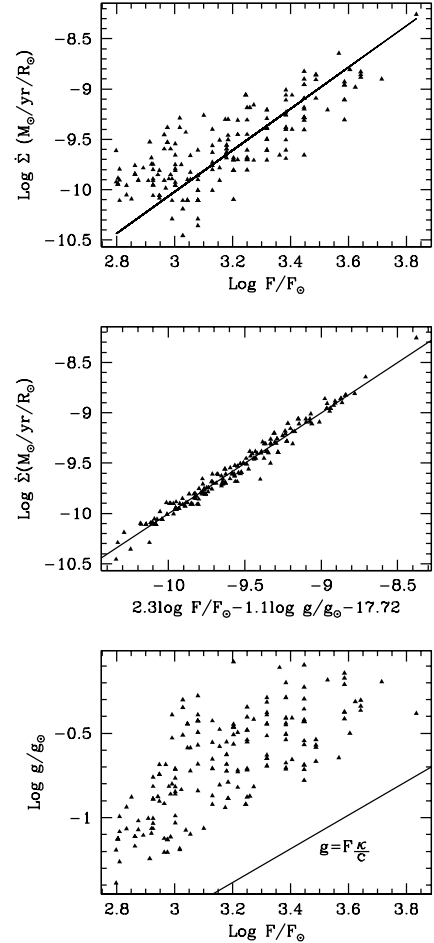}
\caption{
Upper panel: The relation between the mass loss per unit area
$\Sigmadot$ and the local flux $F$, derived from the measured total
mass loss, luminosity, and radius for 201 O stars tabulated by Howarth
\& Prinja (1989). The solid line marks the $\Sigmadot(F)$ relation
used for AGN AD.
Middle panel: The relation between $\Sigmadot$ and both $F$ and
$g$. This rather tight relation follows the predicted CAK relation
between $\Mdot$ and both $L_*$ and $M_*$, ignoring the $1-L/L_{\rm
  Edd}$ term in the CAK solution. The solid line marks the
$\Sigmadot(F,g)$ relation used for AGN AD. Although this relation is
significantly tighter than the $\Sigmadot(F)$ relation derived in the
upper panel, this relation is expected to break in stars at the
Eddington limit, which produces the same local conditions as in AD
atmosphere. Also, the value of $g$ in the standard AD solution assumes
only pure electron scattering opacity, an assumption which breaks down
in AGN AD.
Lower panel: The relation between the surface gravity and the local
flux for the 201 O stars. The solid line shows relation in an
atmosphere supported by radiation pressure, with an electron
scattering opacity, as expected in the inner part of an AGN AD. In O
star atmospheres the radiation force induced by electron scattering
supports up to a half of the local gravity.}
\end{figure}

\section{Analytic estimate of $T_{\rm max}$}

Below we derive the integrated AD wind
\begin{equation}
\Mdot_{\rm wind}(R)=\int_{\infty}^{R}4\pi R\Sigmadot dR .
\end{equation}
based on the relation derived above for $\Sigmadot$ (eqs. 2 \& 3). 
We find the radius 
$r_{\rm eq}$ where $\Mdot_{\rm wind}=\Mdot$, and the thin disc solution likely
terminates. We then find the
local blackbody surface temperature $T_{\rm max}$ at $r_{\rm eq}$, i.e. the hottest
temperature for the thin disc solution.

The flux emitted per unit area from the surface of a thin AD (Shakura \& Sunyaev 1973, hereafter SS73) is 
\begin{equation}
 F=\frac{3}{8\pi}\frac{G\Mdot \Mbh}{R^3}f(R,M,a_*)
\end{equation}
where $R$ is the radius, and $f(R,M,a_*)$ is a dimensionless factor
set by the inner boundary condition, 
and the relativistic effects (Novikov \& Thorne 1973; Riffert \& Herold 1995),
and is of order unity at radii a factor of few larger than the inner boundary.
We assume $f(R,M,a_*)=1$ below. We use the dimensionless radius, $r\equiv R/R_g$,
where $R_g\equiv G\Mbh/c^2$, which gives
\begin{equation}
 F=\frac{3c^6}{8\pi G^2}\frac{\Mdot}{\Mbh^2 r^3} ,
\end{equation}
or equivalently
\begin{equation}
 F/\Fsun=5\times 10^8\Mdott m_8^{-2}r^{-3} ,
\end{equation}
using the relations $M=10^8m_8~\Msun$, $\Mdot=\Mdott~ \Msun~yr^{-1}$,
and a solar flux $\Fsun=6.3\times 10^{10}~{\rm erg}~{\rm s}^{-1}~{\rm cm}^{-2}$.

\subsection{Derivation for $\Sigmadot(F)$}

We now use the AD expression for the local $F$ to derive the expected local $\Sigmadot(F)$
in AD. For convenience,
we express $R$ in equation (5) in solar radii, $R'=R/\Rsun$ where $\Rsun=6.93\times 10^{10}$~cm,
or equivalently $R'=213m_8r$. Thus, equation (5) can be expressed as
\begin{equation}
\Mdot_{\rm wind}(r)=5.69\times 10^5 m_8^2\int_{\infty}^{r} r\Sigmadot(F) dr .
\end{equation}
We now insert equation (8) into equation (2), and get an expression for the local disc $\Sigmadot$
\begin{equation}
\Sigmadot=0.673\Mdott^{1.9}m_8^{-3.8}r^{-5.7} ,\label{eq:sigma1}
\end{equation}
which implies a sharp rise in the local mass loss towards the center.
The integrated mass loss is then
\begin{equation}
\Mdot_{\rm wind}(r)=4.7\times 10^5 m_8^{-1.8}\Mdott^{1.9}r^{-3.7} .
\end{equation}
Thus, $\Mdot_{\rm wind}(r)=\Mdot$ at
\begin{equation}
r_{\rm eq}=34.1m_8^{-0.48}\Mdott^{0.24} ,
\end{equation}
which forms the effective inner boundary of the thin disc solution. Note that
the wind is sharply confined towards $r_{\rm eq}$, as 50\% of $\Mdot_{\rm wind}$
is launched inside $1.2r_{\rm eq}$ and 92\% inside $2r_{\rm eq}$.
The surface effective temperature of a thin AD is (from equation 8)
\begin{equation}
T_{\rm eff}=8.6\times 10^5\Mdott^{1/4}m_8^{-1/2}r^{-3/4}~{\rm K} .
\end{equation}
Thus, the AD temperature at $r_{\rm eq}$ is
\begin{equation}
T_{\rm max}=6.1\times 10^4\Mdott^{0.07}m_8^{-0.14}~{\rm K} .
\end{equation}
Assuming an accretion efficiency of 10\%, the bolometric luminosity is
$L=5.67\times 10^{45}\Mdott$, and $L$ in Eddington luminosity units is
$\mdot=0.44\Mdott m_8^{-1}$. The above expression is equivalent to
\begin{equation}
T_{\rm max}=6.5\times 10^4(\mdot/m_8)^{0.07}~{\rm K} ,
\end{equation}
compared to the $T_{\rm eff}\propto (\mdot/m_8)^{0.25}$ dependence in the SS73 solution
(from equation 13 with $\Mdott$ replaced by $m_8\mdot$).
This simplistic derivation yields that line driven winds from thin
accretion discs in AGN produce an inner boundary with a maximum
temperature of $\sim (5-6)\times10^4$~K, with a weak dependence 
on $\mdot$ and $m_8$. A change by a factor of $10^4$  in 
$\mdot/m_8$ changes $T_{\rm max}$ by only a factor of two.
The wind truncation then explains both the observed position of the UV peak, and
its uniformity, with no free parameters.

\subsection{Derivation for $\Sigmadot(F,g)$}

Below we repeat the above derivation using the above expression
for $\Sigmadot(F,g)$ (equation 3).

We first need to derive the vertical component of gravity, $g$, 
at the disc surface.
The inner AD is supported by radiation pressure, where
the source of opacity is assumed to be dominated by electron scattering.
Thus, $g$ in hydrostatic equilibrium is
\begin{equation}
g=F\kappa_{\rm es}/c ,
\end{equation}
where $\kappa_{\rm es}=0.34~{\rm g}~{\rm cm}^{-2}$ is the electron scattering 
opacity of fully ionized gas. Or, in dimensionless units
\begin{equation}
g/\gsun= 2.6\times 10^{-5}F/\Fsun ,
\end{equation}
where $\gsun=2.74\times 10^{4}$~cm~s$^{-2}$ on the solar surface.
Thus, since $g$ is set by $F$, equation (3) can be rewritten in the
form
\begin{equation}
 \log \Sigmadot =1.21\log F/\Fsun-12.63 ,
\end{equation}
i.e. a weaker dependence on $F$, compared to the $\Sigmadot(F)$
relation (equation 2). Inserting the AD expression for $F$ (equation 8), into the 
above expression yields,
\begin{equation}
\Sigmadot=7.87\times 10^{-3}\Mdott^{1.21}m_8^{-2.42}r^{-3.63} ,\label{eq:sigma2}
\end{equation}
and following the integration we get
\begin{equation}
\Mdot_{\rm wind}(r)=2.75\times 10^3 m_8^{-0.42}\Mdott^{1.21}r^{-1.63} .
\end{equation}
We thus get a significantly weaker rise in $\Mdot_{\rm wind}$ with decreasing 
$r$, compared to the one derived from the $\Sigmadot(F)$ relation (equation 11).
We now get
\begin{equation}
r_{\rm eq}=129m_8^{-0.26}\Mdott^{0.13} .
\end{equation}
Note that in this case the wind is somewhat less sharply confined towards 
$r_{\rm eq}$, compared to the $\Sigmadot(F)$ case. Here,
50\% of $\Mdot_{\rm wind}$ is launched inside $1.5r_{\rm eq}$ and 92\% 
inside $4.7r_{\rm eq}$, compared to  $1.2r_{\rm eq}$ and $2r_{\rm eq}$
in the $\Sigmadot(F)$ case. We also get 
\begin{equation}
T_{\rm max}=2.25\times 10^4\Mdott^{0.153}m_8^{-0.305}~{\rm K} .
\end{equation}
or
\begin{equation}
T_{\rm max}=1.99\times 10^4(\mdot/m_8)^{0.153}~{\rm K} ,
\end{equation}
which is steeper than the $(\mdot/m_8)^{0.07}$ dependence derived for the $\Sigmadot(F)$ solution
(equation 15), but is still flatter than the SS73 dependence of $(\mdot/m_8)^{0.25}$. The value of
$T_{\rm max}$ here is lower than for the $\Sigmadot(F)$ solution.

The wind flux is inversely correlated with $g$ (equation 3). The value of $g$ in hydrostatic
equilibrium depends linearly
on the gas opacity (equation 16). The electron scattering opacity used here is the minimal
opacity for ionized gas. The additional contribution from line opacity increases $g$,
and thus decreases $\Sigmadot$. As a result, the disc may extend further inwards, 
and thus reach a higher temperature than derived above (eqs. 22, 23). 

\section{The Navier-Stokes equations for an AD with mass loss}
The above estimates suggest that line driven winds from AGN AD prevent
the formation of the hot inner AD regions with $T>10^5$~K, which may explain the uniformity
of the FUV SED of AGN. However, these estimate are rather crude, as the expression used for $F$
(equation  8) ignores the reduction in $\Mdot$ due to the wind mass loss. Below we derive the AD  structure, based on the
mass, momentum and energy continuity equations, including a wind mass loss term.
We then calculate the revised AD SED, first using the local blackbody approximation, and 
then using the stellar atmospheric solution code TLUSTY \citep{hl95,hub00}.

The derivation below is for a viscous flow, described by the
Navier-Stokes equations.  We use cylindrical coordinates, $R,z, \phi$,
and assume axial symmetry (no $\phi$ dependence).  We further assume
the $R$ and $z$ solutions are separable, which is likely valid in the
thin disc approximation, and we solve for the radial dependence
only. The solution below yields the radial dependence of the
vertically integrated viscous torque $W_{R\phi}(R)$, which is required
in order to get a steady state solution. This quantity, together with
$\Omega(R)$ - the angular velocity radial dependence, uniquely determine $F(R)$.  The
physical origin of $W_{R\phi}(R)$ is an open question, heuristically
addressed by the $\alpha$ disc model (SS73).  In ionized accretion
discs, it is now widely believed  that angular momentum transport is 
provided by magnetorotational turbulence \citep{bh98}, which when averaged 
over time and the vertical extent of the
disc, seems to be reasonably approximated by an $\alpha$-disc solution
\citep{1999ApJ...521..650B}.  If the disc is thin, only the surface
density, $\Sigma$, and the vertical structure of the disc, depend on
the accretion stress mechanism. The expression for
$F(R)$ and the derived SED, in the local blackbody approximation, are
independent of the nature of the angular momentum transport
mechanism. A relation for the accretion stress is required to
derive a detailed model of the AD vertical structure, which can then
be used to derive the local $\Sigmadot$ from first
principles, as done by CAK for O stars.  Although the $\alpha$ disc
model allows to solve the vertical disc structure, it is just a
convenient way to parametrize our ignorance, and is far from being a
first principles solution. Below we circumvent this difficulty by
adopting the stellar $\Sigmadot$ as
described above.

\subsection{Derivation}
\label{derivation}

The time-dependent AD equations can be derived by
formulating the Navier-Stokes equations in cylindrical coordinates with
vertical averaging \citep[see e.g.][]{1999ApJ...521..650B}.  When
the disc is sufficiently thin, the radial momentum equation is to
lowest order simply a balance between rotational terms and gravity,
with a slow radial inflow due to the stress.  The
gravitational potential then determines the rotation rate $\Omega(R)$,
which is Keplerian for a point source. Balbus \& Papaploizou (1999,
see also Blaes 2004) show that with appropriate
averaging, stresses arising from magnetorotational turbulence yield
(to lowest order in $H/R$) essentially identical relations to the viscous
relations when written in terms of the the vertically integrated
stress $W_{R\phi}$\footnote{Note that our definition of $W_{R\phi}$
  differs from \citet{1999ApJ...521..650B} by a factor of $\Sigma$.}

We now generalize these equations to include mass outflow from the
disc surface.  We find conservation equations for the mass:
\begin{eqnarray}
\frac{\partial \Sigma}{\partial t}+\frac{1}{R}\frac{\partial}{\partial R}
\left( \Sigma v_R R\right)=-2 F_M,\label{eq:mass}
\end{eqnarray}
angular momentum:
\begin{eqnarray}
\frac{\partial}{\partial t}\left( \Sigma R^2 \Omega \right)+
\frac{1}{R} \frac{\partial}{\partial R}\left( R^3 \Omega \Sigma v_R + R^2 W_{R\phi} 
\right)= 
-2 R^2\Omega F_M,\label{eq:angm}
\end{eqnarray}
and energy:
\begin{eqnarray}
F=-\frac{W_{R\phi} R}{2}\frac{\partial \Omega}{\partial R}
 - \epsilon \frac{\Omega^2 R^2}{2} F_M.
\label{eq:energy}
\end{eqnarray}
Here, $\Sigma$ is the surface density, $v_R$ is the radial velocity, and
$F$ is the radiative flux from one side of the disc.  

These are identical to eqs. (26), (27) and (46) of
\citet{1999ApJ...521..650B} except for the appearance of fluxes of
mass $F_M$, angular momentum, and energy due to the outflow.  The
surface terms no longer vanish in the vertical integration, giving
$F_M \simeq \rho |v_z|$, corresponding to a vertical momentum flux.
This mass flux carries away an angular momentum flux proportional to
the specific angular momentum of the material at its launching radius.
Note that we have not attempted to account for an additional torque of
the wind on the disc that might arise if e.g. the wind and disc are
magnetically coupled.  In some cases, the torque may be plausibly
absorbed into $W_{R\phi}$ if there is associated dissipation that can
be modelled as in equation  (\ref{eq:energy}).  In general, this depends on
the details of the torque mechanism \citep[see
  e.g.][]{1999ApJ...521..650B}.

The second term on the right hand side of equation  (\ref{eq:energy})
accounts for possible work done by the radiation field in unbinding
the outflow. Since the vertical component of gravity continues to
increase (initially linearly), the radiation field will do work
against gravity launching and accelerating any unbound material.  We
introduce a parameter $\epsilon$, which corresponds to the fraction of
the gravitational binding energy transferred from the radiation to the
outflow, once the outflow leaves the thin disc.  Hence, $\epsilon =1$
corresponds to a flow where all material removed from the thin disc
reaches the local escape velocity.

If $\epsilon$ is treated as a constant, this prescription implies that
the mass launched from an annulus at radius $R$ is accelerated locally.
Of course, this is generally not
correct and more sophisticated calculations and numerical simulations
\citep[e.g.][]{1995ApJ...451..498M,2000ApJ...543..686P} show that most
of the outflow is accelerated above the surface of the disc, and 
predominately by radiation from regions
interior to its launching radius. A realistic outflow model requires a global 
numerical simulation, and is beyond the
scope of this paper.  In the following, we simply adopt
equation  (\ref{eq:energy}) as a useful heuristic which aids in the
discussion of global energy conservation.  We offer some discussion of
the global aspects of the AD and outflow in Section 7.

We now focus on time steady solutions of accretion discs with Keplerian
rotation $\Omega_K^2=GM/R^3$. We assume $F_M = \Sigmadot$ and
adopt the standard definition of mass accretion rate  $\dot{M}=-2\pi \Sigma R v_R$.
Then equations (\ref{eq:mass}) and (\ref{eq:angm}) become
\begin{eqnarray}
\frac{\partial \dot{M}}{\partial R} =4 \pi R \Sigmadot,\label{eq:massss}
\end{eqnarray}
and
\begin{eqnarray}
\frac{\partial}{\partial R}\left(\dot{M} \Omega_K R^2 - 2 \pi W_{R\phi} R^2 
\right)=4 \pi R^3\Omega_K \Sigmadot.\label{eq:angmss}
\end{eqnarray}

\subsection{The no wind solution}
In the standard thin disc with no wind $\Sigmadot=0$ and 
\begin{eqnarray}
\frac{\Mdot \Omega_K R^2}{2\pi} - W_{R\phi} R^2=C_1,\nonumber
\end{eqnarray}
where $C_1$ is a constant independent of $R$.  Assuming $W_{R\phi}=0$ at $R_{\rm in}$ gives
\begin{eqnarray}
W_{R\phi}=\frac{\dot{M} \Omega_K}{2\pi}\left[1-\left(\frac{R_{\rm in}}{R}\right)^{1/2} \right],\label{eq:wss73}
\end{eqnarray}
which reduces to the standard SS73 expression for $F$ when inserted
into equation  (\ref{eq:energy}).

\subsection{The disc + wind equations}

For a Keplerian disc with mass loss, eqs. (\ref{eq:energy}),
(\ref{eq:massss}), (\ref{eq:angmss}) provide a system of coupled
partial differential equations, since $\Sigmadot$ 
depends on $F$, which is computed 
using equation (\ref{eq:energy}). For the form of $\Sigmadot$ 
given in eqs. (2) and (3), there
is no simple analytical solution, and these equations must be
integrated numerically.

\subsubsection{An example for an analytic solution}
For illustrative purposes we derive below a simple analytic solution
for $F(R)$ for a Keplerian disc with a given analytic expression for
$\Mdot_{\rm wind}(R)$ and $\epsilon = 0$.  This provides some insight on the effect of a wind on the
AD SED.  

Let us assume a simple power law expression of $\Mdot_{\rm wind}(R)=\Mdot_0(R/R_0)^{\alpha}$,
where $\Mdot_0$ is the accretion rate at $r=\infty$, $\alpha<0$, and the disc terminates due to the wind at 
$R_0=r_{\rm eq}R_g$. This then gives
\begin{equation}
\Mdot(R)=\Mdot_0[1- (\frac{R}{R_0})^{\alpha}] .
\end{equation}

Using equation  (\ref{eq:wss73}) with a Keplerian disc, gives
\begin{equation}
W_{R\phi}(R)R^2=\frac{\sqrt{GM}}{4\pi}\Mdot_0\int_{R_0}^R R^{-1/2}[1- (\frac{R}{R_0})^{\alpha}]dR .
\end{equation}
The lower integration limit gives the boundary condition $W_{R\phi}(R_0)=0$. 
We get (for $\alpha\neq -1/2$)
\begin{equation}
W_{R\phi}(R)=\sqrt{\frac{GM}{R^3}}\frac{\Mdot_0}{2\pi}[1- \frac{2\alpha}{1+2\alpha}(\frac{R_0}{R})^{1/2}
-\frac{1}{1+2\alpha}(\frac{R}{R_0})^{\alpha}] .
\end{equation}
The local flux is then 
\begin{equation}
F(R)=\frac{3}{8\pi}\frac{GM\Mdot_0}{R^3}[1- \frac{2\alpha}{1+2\alpha}(\frac{R_0}{R})^{1/2}
-\frac{1}{1+2\alpha}(\frac{R}{R_0})^{\alpha}] ,
\end{equation}
which gives the no wind solution
\begin{equation}
F(R)=\frac{3}{8\pi}\frac{GM\Mdot_0}{R^3}[1-(\frac{R_0}{R})^{1/2}] ,
\end{equation}
for $\alpha\rightarrow -\infty$, as expected. Note that inserting $\Mdot(R)$ from equation (30) instead
of $\Mdot_0$ into equation (34), to get the effect of mass loss, is not a valid solution, and
yields a higher value
for $F(R)$ compared to equation (30), e.g. by 50\% for $R/R_0=2$ for $\alpha=-2$.

\section{Results}
\label{results}


Below we consider models which parametrize the mass flux with either
$\Sigmadot(F)$ or $\Sigmadot(F,g)$. For each model we consider cases
with both $\epsilon=0$ and $\epsilon = 1$ in equation
(\ref{eq:energy}).  The models with $\epsilon = 0$ do not account 
for the energy lost in unbinding the flow while the set of models with $\epsilon
\ge 1$ assumes that all $\dot{M}_{\rm wind}$ becomes unbound and
escapes the system. As we will see, models with lower $M$  and
high $\mdot$ can lead
to very large implied mass outflow rates with $\dot{M}_{\rm wind}
\simeq \dot{M}_{\rm out}$.  In this regime, the fraction of
$\dot{M}_{\rm out}$ that is removed from the disc is very sensitive
to our choice of $\epsilon$, but we shall see that the
derived radiative flux from the disc is rather insensitive to this assumption.


We present the derived $T_{\rm eff}(r) \equiv (F/\sigma_{\rm
  B})^{1/4}$ for the disc+wind solution, and the associated SED (note
that $r=R/R_g$). We show the significantly reduced dependence of
$\nu_{\rm peak}$ on $\mdot/M$, as expected from the simplified
analytic derivation above (eqs. 15, 23).  In the absence of a wind,
the inner disc radius is often assumed to correspond to the innermost
stable circular orbit $r_{\rm ISCO}$, which in turn is set by the
black hole spin $a_*$. In the absence of a wind, higher $a_*$ AD
spectra are harder. Below we show that since generally $r_{\rm
  eq}>>r_{\rm ISCO}$, the value of $a_*$ has no effect on the observed
SED, as the innermost thin disc region is gone with the wind.  We also
show below that sufficiently cold discs are not affected by the wind,
and the SED of objects with low enough $\mdot/M$ values should be well
fit by the standard disc solution.

\subsection{The numerical solution}

In the appendix, we describe the general relativistic generalizations
of eqs.  (\ref{eq:energy}) and (\ref{eq:angmss}), which now depend
explicitly on $a_*$.  Eqs.  (\ref{eq:angmf}) and (\ref{eq:energyf}) 
still form a closed set of two coupled equations that need to be
numerically integrated, with boundary conditions for both $\dot{M}$
and $W_{R\phi}$.  If we set our boundary condition at $r_{\rm in}$ as
in SS73, $\dot{M}_{\rm in}$ is a parameter of the problem.  We follow
SS73 by setting $W_{R\phi}(r_{\rm in})=0$ and integrate outward.
Alternatively, we could start at a large radius where $T_{\rm eff}$ is
low, $\Sigmadot=0$ and $\dot{M}=\dot{M}_{\rm out}$
parametrizes the disc model.  However, there is some ambiguity in the
specification of $W_{R\phi}(r_{\rm out})$ in this case.  This is
problematic for models where $\dot{M}_{\rm wind}$ becomes a
substantial fraction of $\dot{M}(r_{\rm out})$, as the integrated mass
loss and its radial profile can be quite sensitive to the precise
choice of $W_{R\phi}(r_{\rm out})$.  Therefore, we focus our attention
on solutions that start from $r_{\rm in}$ and integrate outward.


Since our model presumes that $W_{R\phi}$ is generated by {\em local}
stresses within the disc, there is no reason to expect that
$W_{R\phi}$, set by local conditions at $r \gg r_{\rm in}$, happens to
have the exact value required to match on to a stationary solution with
$W_{R\phi}(r_{\rm in})=0$.  In a disc with no outflow, matter should
have time to diffusively adjust as it slowly spirals in, since the
viscous time decreases inward.  Hence, it should generally follow the
equilibrium solution specified by the inner torque assumption, unless
an instability in the flow is present (see e.g. Shakura \& Sunyaev
1976).  However, it is less clear whether the same argument applies to
the models considered here when there is significant mass outflow.  It
is possible that such flows show substantial time variability, but
we only explore steady state models here.

\begin{figure}
\includegraphics[width=84mm]{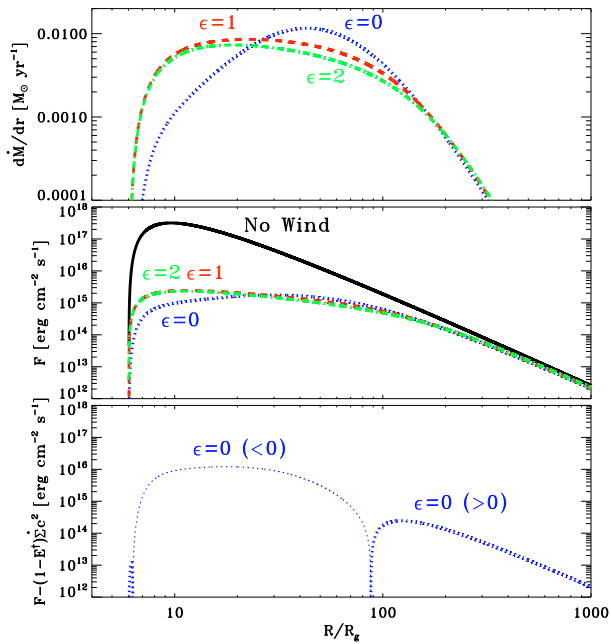}
\caption{ 
Upper panel: The implied mass-loss rate versus $R$ for models with
$\epsilon=0$ (blue, dotted), 1 (red, dashed) and 2 (green,
dot-dashed).  All models correspond to $M/\Msun=10^7$, $a_*=0$ and
 $\mdot=2$ at a large radius.  
Mass loss is modelled using the $\Sigmadot(F)$ prescription.
The integrated mass loss decreases as $\epsilon$ increases, with
the largest outflow coming from the $\epsilon=0$ model, for which
a larger fraction of the outflow occurs at a larger radius.
Middle panel: The flux corresponding to the model shown in the upper
panel and a model with no outflow (solid black).  The larger fraction
of energy that is lost in unbinding the flow for larger $\epsilon$
is offset by lower implied outflow rates and the flux profiles for the
models with mass loss are all very similar.
Bottom panel: The difference between the radiative flux and the energy
required to unbind the implied outflow for the model with $\epsilon=0$.
The thin dotted curve shows where this quantity is negative.  The
negative value implies that mass loss is so great that there is insufficient
energy to unbind it.  Hence, most of the mass lost from the thin disk,
if $\epsilon=0$, must ultimately accrete, possibly in the form of a much hotter and 
geometrically thick flow.}
\label{f:eps_comp}
\end{figure}

Figure \ref{f:eps_comp} shows a comparison of models with $\epsilon
=0$, 1, and 2 for a $10^7 \Msun$, $a_*=0$ black hole with mass loss
parametrized by $\Sigmadot(F)$ and $\dot{m}_{\rm out}=2$.  The model
with $\epsilon=1$ launches a wind where all material just reaches its
escape velocity, while model with $\epsilon=2$ reaches infinity with a
kinetic energy which equals the binding energy at its launching radius.
The top panel shows that the overall mass loss is larger for the
$\epsilon=0$ model, and peaks at a somewhat larger radius than the
$\epsilon \ge 1$ models.  The $\epsilon=1$ and 2 profiles for
$\dot{M}_{\rm wind}(r)$ are very similar and the outflow is more
broadly distributed in radius.  The middle panel shows the
corresponding $F(r)$ for the models in the upper panel, evaluated using
equation (\ref{eq:energyinf}).  The outflow models all yield
significantly lower values than the no mass loss model.  
The strong sensitivity of $\Sigmadot$ on $F$ (equations 2 and 3) 
serves as a thermostatic effect on the maximal possible $F$ value
in all models.  In the $\epsilon \ge 1$ models,
this happens primarily through the second term on right-hand-side of
equation (\ref{eq:energyinf}) so that the larger fraction of energy
that is lost in unbinding the flow is offset by
lower implied outflow rates. In the $\epsilon=0$ case, the cap on $F$
only occurs through a reduction of $W_{r\phi}$, which depends less
directly on $\Sigmadot$ through equation (\ref{eq:angmcons}). Due to
the strong similarity of $F(r)$ in the models with $\epsilon=1$ and 2, we only
show the $\epsilon=0$ and 1 cases as representative examples in
subsequent plots.

The $\epsilon=0$ case assumes no kinetic energy is taken by the outflow.
Can the outflow still gain enough energy by intercepting enough
of $F(r)$, say in the form of a radiation pressure driven wind, to
produce a wind which escapes to infinity? The bottom panel explores 
this question by showing a modified $F(r)$, for the $\epsilon=0$ model,
when we subtract from the original value the binding energy of the
mass lost.  We compute this by
subtracting off the flux of kinetic energy $(1-E^\dagger)\Sigmadot
c^2$, which is required to (just) unbind the outflow. This value
becomes negative at $r<90$.  Hence, there is
insufficient energy in the radiation field to unbind the implied
outflow. So, if there is no physical mechanism which can provide 
$\epsilon>0$, i.e. a mechanism which can convert directly some of the 
local dissipated energy in the disk into kinetic energy of the mass lost,
 then a large fraction of the implied $\dot{M}_{\rm wind}$ must
ultimately form a ``failed wind'' and accrete.  We discuss the
observational implications of such failed winds in section 7.2.

\begin{figure}
\includegraphics[width=84mm]{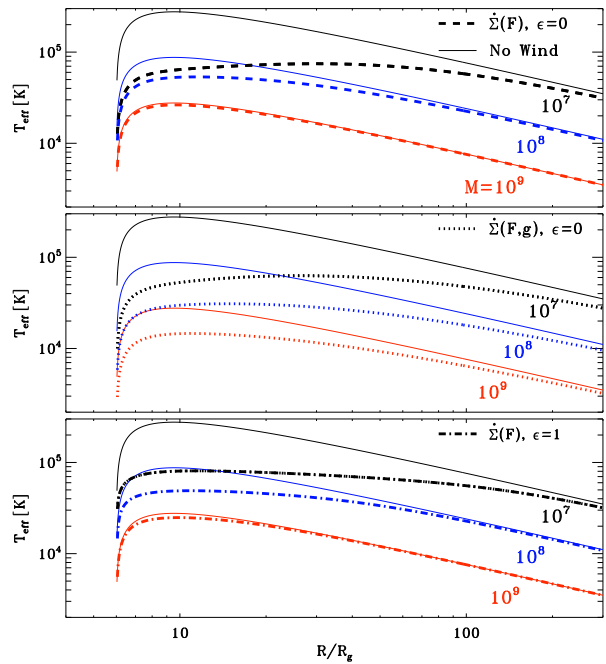}
\caption{ 
Upper panel: The $T_{\rm eff}$ versus $R$ for models with
$M/\Msun=10^7$, $10^8$,  and $10^9$.  All models
correspond to $a_*=0$ and $\Mdott=0.94$ at a large radius, which
corresponds to $\mdot=2$, 0.2, and 0.02, respectively.  The solid
curves are standard AD models with no mass loss while the
dashed curves show models with mass loss given by $\Sigmadot(F)$ with
$\epsilon=0$. Note the reduced difference in $T_{\rm max}$ between the
$M/\Msun=10^7$ and $10^8$ mass loss models, compared to the standard
solution. The AD in the $M/\Msun=10^9$ model is too cold to produce
significant mass loss, and $T_{\rm eff}$ profile remains the same.
Middle panel: Same as the upper panel but the dotted curves
correspond to $\Sigmadot(F,g)$ with $\epsilon =0$.  This mass loss
term yields a weaker $R$ dependence of $\Mdot$, and thus a more
gradual rise at smaller $R$ in the deviation of $T_{\rm eff}$ from the
standard solution.
Bottom panel: Same as the upper panel, but the dot-dashed
curves correspond to $\Sigmadot(F)$ with $\epsilon=1$.  In
this case the $T_{\rm eff}$ continues to increase very mildly
as radius decreases, rather than decreasing slightly as in the $\epsilon=0$
case.}
\label{f:mass_teff}
\end{figure}

\begin{figure}
\includegraphics[width=84mm]{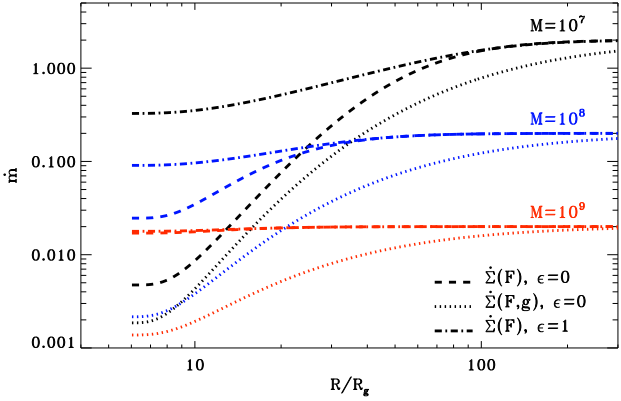}
\caption{ 
The local $\mdot$ versus $R$ for the models shown in
Fig. \ref{f:mass_teff}.  The implied drop in $\mdot$ towards $R_{\rm
  ISCO}$ becomes larger with decreasing $M$, as the AD gets hotter,
and can reach a factor of 1000 in the most extreme model,
corresponding to $M=10^7 \Msun$, $\Sigmadot(F,g)$, and $\epsilon=0$.
The models with $\epsilon=1$ generically produce less implied mass
loss, with most extreme case ($M=10^7 \Msun$) having $\dot{M}_{\rm
  out} \sim 6 \dot{M}_{\rm in}$.  The $M/\Msun=10^9$ and
$\Sigmadot(F)$ models show no significant mass loss for either value
of $\epsilon$.}
\label{f:mass_mdot}
\end{figure}

Figure \ref{f:mass_teff} presents numerical solutions for $T_{\rm
  eff}(r)$. The results are presented for the two $\Sigmadot$
relations, and for different $M$ values.  The top  panel presents
the $\Sigmadot(F)$, $\epsilon=0$ AD model solution for $T_{\rm
  eff}(r)$, for $M=10^7, 10^8, 10^9 M_\odot$. All models correspond to
$a_*=0$ and $\Mdott=0.94$ at $r \gg r_{\rm in}$, which corresponds to
$\mdot=2$, 0.2, and 0.02, respectively.  For comparison we also show
the corresponding standard SS73 solution for $T_{\rm eff}(r)$ with no
mass loss, i.e. a constant $\mdot$ ($=\mdot(r_{\rm out})$).  Note the
similar $T_{\rm max}$ in the $M=10^7M_\odot$ and $10^8 M_\odot$
models, with a ratio of 1.4, in contrast with the ratio of
$3.16$ for the SS73 solution (equation 13). The simplified
analytic solution ratio (equation  14) of $10^{0.14}=1.38$, is remarkably
close to the numerical solution ratio of 1.4.  The absolute values of $T_{\rm
  max}$ in the analytic solution, $7.2\times 10^4$~K for
$M=10^7M_\odot$ is also very close to the numerical solution value of
$7.5\times 10^4$~K.  The $M=10^9M_\odot$ model is cold enough to
suppress $\Sigmadot(F)$, so that $\mdot_{\rm wind}(r)/\mdot\ll 1$, and
the solution overlaps the SS73 no wind solution.

The middle panel shows the numerical solution using $\Sigmadot(F,g)$
and $\epsilon=0$ for the same parameters as in the upper panels. The
mass loss is more pronounced, and its
rise towards the center is more gradual, as expected from the
simplified analytic solution (equation  20 vs. equation  11). The wind remains
significant also for the $M=10^9M_\odot$ models.  The ratios of
$T_{\rm max}$ from the $M=10^7M_\odot$ and $10^8 M_\odot$ models are,
2.03, somewhat larger than with the $\Sigmadot(F)$ relation.

The bottom panel shows a model with $\Sigmadot(F)$ and
$\epsilon=1$. Although the implementation of the outflow differs
significantly from the top panel, the thermostatic effect is still
apparent.  In this case the ratio of $T_{\rm max}$ for the
$M=10^7M_\odot$ and $10^8 M_\odot$ models is 1.65, compared to 1.4 
for the $\epsilon=0$ model above, and significantly less than the SS73 model.
As above, there is very little mass lost for the $M=10^9M_\odot$
model.  Models with $\Sigmadot(F,g)$ and $\epsilon=1$ (not shown)
have $T_{\rm eff}(r)$ profiles qualitatively similar to those with
$\Sigmadot(F,g)$ and $\epsilon=0$.

Figure \ref{f:mass_mdot} shows the implied $\dot{m}(r)$ profiles for
the models in Figure \ref{f:mass_teff}.  In the simplified analytic
solution the disc truncates at $r_{\rm eq}$, where $\dot{m}_{\rm
  wind}(r_{\rm eq})=\dot{m}(r_{\rm out})$. In the numerical solutions,
the drop in $\dot{m}(r)$ towards the center suppresses the rise in
$T_{\rm eff}(r)$, and thus reduces a further rise in $\dot{m}_{\rm wind}(r)$
towards the center. This negative feedback prevents $\dot{m}_{\rm
  wind}(r)$ from ever reaching $\dot{m}(r_{\rm out})$, and the disc
always (nominally) extends down to $r_{\rm ISCO}$. However, the
implied total mass loss can be very large.  The $\Sigmadot(F,g)$ model
with $\epsilon=0$ yields $\dot{m}(r_{\rm in})/\dot{m}(r_{\rm
  out})=0.00092$ for $M=10^7M_\odot$, and the $\Sigmadot(F)$ model
with $\epsilon=0$ yields $\dot{m}(r_{\rm in})/\dot{m}(r_{\rm
  out})=0.0024$. Thus, although formally the thin disc extends down to
$r_{\rm ISCO}$, it effectively terminates at a larger radius in these
cases.  For example, $\dot{m}(r)/\dot{m}(r_{\rm out})=0.5$ at $r=62$
for the $\Sigmadot(F)$ $M=10^7M_\odot$ model, and at $r=20$
for the $M=10^8M_\odot$ model (the simplified analytic solution,
equation 12, gives truncation radii of $r=85$ and 28 respectively).  Inside these
transition radii, the implied $\dot{m}(r)$ profiles should be regarded
with suspicion, given a possible feedback of the mass loss on the thin disk
solution. Obtaining reliable profiles in this region probably
requires a more sophisticated disc model.

The coldest models with $M=10^9M_\odot$ has no significant mass loss
for models employing the $\Sigmadot(F)$ relation. This is consistent
with the $T_{\rm eff}(r)$ solution in the upper and lower panels of
Figure \ref{f:mass_teff}, which matches the no wind SS73 solution, as
$\dot{m}(r)$ remains effectively constant. The $\Sigmadot(F,g)$
relation yields a more gradual drop in $\dot{m}(r)$ towards the center
for all $M$, but with a significantly larger amplitude, with
significant wind also for $M=10^9M_\odot$.

\begin{figure}
\includegraphics[width=84mm]{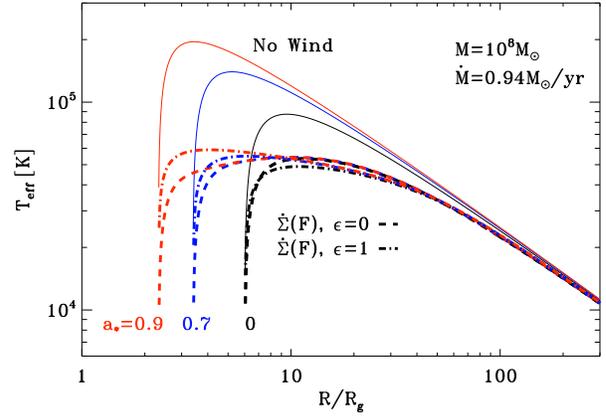}
\caption{ 
The $T_{\rm eff}(r)$ solution for models with $a_*=0$, $0.7$
and $0.9$.  All models correspond to $M=10^8 \Msun$ and
have $\Mdott=0.94$ at a large radius, which corresponds to $\mdot=0.2$,
0.36, and 0.54, respectively.  The solid curves denote standard accretion
disc (SS73) models with no mass loss.  The dashed and dot-dashed
curves denote models with mass loss given by $\Sigmadot(F)$ for
$\epsilon=0$ and $\epsilon=1$, respectively. In contrast with the SS73
models, all $\epsilon=0$ models reach the same $T_{\rm max}$. This
reflects the thermostatic effect of $\Sigmadot$, which sets a cap on
$T_{\rm max}$, and produces a nearly isothermal AD at $r\sim 2-20$.
For the $\epsilon=1$ models, $T_{\rm eff}$ continues to rise to small
$R$, but much more weakly than in the SS73 models.  Since the emitting area
decreases, these innermost radii contribute little to the disc
integrated luminosity. At larger $R$, which dominate the bolometric
output, the profiles nearly overlap, as in the $\epsilon=0$ models.
\label{f:spin_teff} }
\end{figure}

Figure \ref{f:spin_teff} explores the effect of the black hole spin
$a_*$ on $T_{\rm eff}(r)$, for the models assuming
outflow rates of $\Sigmadot(F)$ and $\epsilon=0$ and 1. The solutions 
of $T_{\rm eff}(r)$ are presented for $a_*=0,
0.7, 0.9$, All models correspond to $M=10^8M_\odot$, and $\Mdott=0.94$
at large $r$, which corresponds to $\mdot=0.2$, 0.36, and 0.54,
respectively. The SS73 solutions are also shown for comparison. In
contrast with the SS73 solution, where $T_{\rm max}$ rises with $a_*$,
as $r_{\rm ISCO}$ gets smaller, in the $\epsilon=0$ case all models reach a
nearly identical
$T_{\rm max}$, which occurs at $r\simeq 10$, well outside the largest
$r_{\rm ISCO}=6$.  The value of $T_{\rm max}$ is well below the SS73
range of values, as shown above in Figure \ref{f:mass_teff}. The AD
extension to smaller $r$ with increasing $a_*$ just produces an
extended inner region with $T\simeq T_{\rm max}$, from $r\simeq
20$ down to $r\simeq r_{\rm ISCO}$. The strong dependence of
$\Sigmadot$ on the local $F$ effectively serves as a local thermostat,
which prevents $F$ from rising above the limiting $T_{\rm max}$ value.
Thus, the wind breaks the tight relation between $T_{\rm max}$ and
$a_*$, which exists in models with no mass loss.

The $\epsilon=1$ models show a slight rise in $T_{\rm eff}(r)$ 
towards smaller $r$, but since the emitting area scales as $r^2$, we
will see that this weak increase has almost no effect on the observed SED.

The results presented in Figures \ref{f:mass_teff} and
\ref{f:spin_teff} clearly show that $T_{\rm max}$, remains well below
$10^5$K.  There is some dependence of $T_{\rm max}$ on the AD
parameters, but the dependence is significantly reduced compared to
the solutions with no winds, particularly for the $\Sigmadot(F)$
relations. The numerical solutions are rather close to the simplified
analytic estimate made in \S 3 for $r_{\rm eq}$ and $T_{\rm max}$.
The qualitative similarities between the models with $\epsilon=0$ and
$\epsilon=1$ suggest the thermostatic effect of the wind could be
quite robust.  The difference between the $\epsilon=0$ and 1 models is
their significantly different $\mdot(r)$ in the hottest models.  
The models utilizing
the $\Sigmadot(F,g)$ outflow prescription, generally provide
too much outflow, leading to discs considerably colder than those observed.

\section{The Derived SED}

\subsection{The local Blackbody models}

\begin{figure}
\includegraphics[width=84mm]{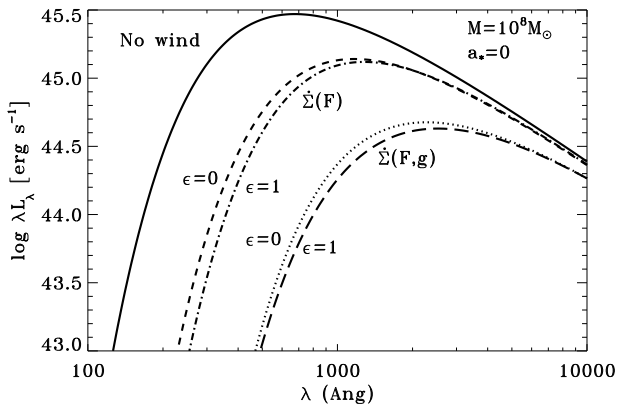}
\caption{
The specific luminosity $\lambda L_\lambda$ versus $\lambda$ for
blackbody emission corresponding to the models with different
$\Sigmadot$ and $\epsilon$, as noted in the plot. All models assume no torque
at the inner boundary, $a_*=0$, $M=10^8 M_\odot$, and $\dot{m}=0.2$ at large 
$R$. The standard model
with no mass loss (solid) peaks at $\lambda \sim 600$\AA, significantly
shorter than typically observed. The $\Sigmadot(F)$ models peak at
$\lambda \sim 1000$\AA, as typically observed. The $\Sigmadot(F, g)$
models, characterized by a stronger and more extended mass loss, peaks at
$\lambda > 2000$\AA, colder than typically observed.}
\label{f:comp_spec_model}
\end{figure}

We now consider the SED predicted by the disc models in the presence
of a mass outflow.  The outflow can modify the spectrum in two primary
ways: by modifying the underlying thin disc solution, as derived
above, and also via its direct emission or reprocessing (absorption
and scattering) of radiation from the underlying disc.  In this work
we focus only on the effects on the underlying thin disc solution,
which may produce the universal turnover at $\lambda<1000$\AA. A more
complete calculation requires modeling the outflow
\citep{1995ApJ...451..498M,2000ApJ...543..686P} to compute the effect
of reprocessed emission \citep[see e.g.][]{2010MNRAS.408.1396S} on the
SED.  Such models require detailed numerical simulations that are
beyond the scope of this work.

We first study the derived SED based on the simple local blackbody SED
calculations, computed directly from the $T_{\rm eff}(r)$ profiles
discussed above.  We later present the results from a more detailed
model which includes radiative transfer and the vertical structure of
the atmosphere. The advantage of the simplified local
blackbody calculation is that the results are insensitive to
assumptions about the vertical dissipation
distribution.  We break the disc up into concentric annuli equally
spaced in $\log r$.  At each radius we compute a blackbody spectrum at
$T_{\rm eff}(r)$, weighting by the emitting area of the annulus and
summing over all radii yields full disc SEDs. For the sake of
simplicity, the effects of relativity on photon propagation are
neglected at this stage, and are included in the following more
detailed atmospheric calculations.

Figure \ref{f:comp_spec_model} presents the SED for the outflow
prescriptions described in \S 5.1.  We use the $T_{\rm eff}(r)$
profiles presented in Figure \ref{f:mass_teff} for $M=10^8 M_\odot$, 
$\dot{m}(r_{\rm out})=0.2$, and $a_*=0$.
The blackbody SED with no outflow peaks at $\sim 600$ \AA, too far in the
UV to be consistent with the observed SEDs of $10^8 M_\odot$ black
holes. The $\epsilon=0$ and 1 $\Sigmadot(F)$ models both 
yield peaks at $\sim 1000$ \AA, as
typically observed.  The $\Sigmadot(F, g)$ models, which predicts more
mass loss, peaks long-ward of $2000$ \AA, too long to be consistent
with observations of most luminous AGN. 

\begin{figure}
\includegraphics[width=84mm]{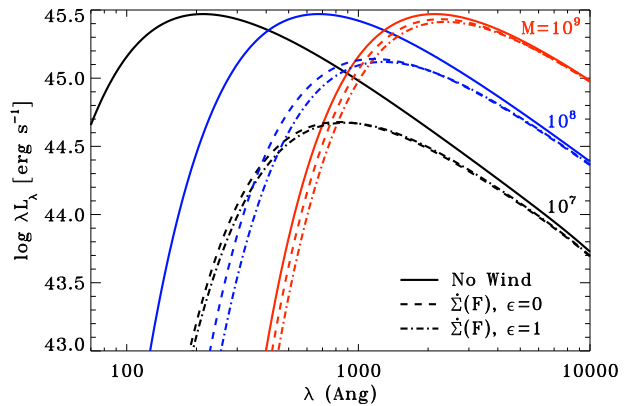}
\caption{
The specific luminosity $\lambda L_\lambda$ versus $\lambda$ for
blackbody emission corresponding to the models presented in Figure
\ref{f:mass_teff} using the $\Sigmadot(F)$ prescription with
$\epsilon=0$ (dashed) and $\epsilon=1$ (dot-dashed).  The peak
emission wavelength decreases with $M$ in all models, but in the
models with outflow this sensitivity to $M$ is significantly
reduced. With these relations, the $10^9 M_\odot$ models are cold
enough to have a negligible wind effect on the SED. The integrated
luminosity drops with $M$ when mass loss is included in the models, as
expected from the $\mdot(r)$ solutions.  There are only very
modest difference between the $\epsilon=0$ and $\epsilon=1$ models,
suggesting the thermostatic effect on the SED is fairly robust.}
\label{f:comp_spec_mass}
\end{figure}

Figure \ref{f:comp_spec_mass} presents the SEDs as a function of $M$. 
It compares the SEDs derived from
the $\Sigmadot(F)$ model for $\epsilon=0$ and $\epsilon=1$ with the
standard SS73 SEDs. As expected from the $T_{\rm eff}(r)$ solutions (Figure 2),
the $M=10^7M_\odot$ and $M=10^8M_\odot$ wind solution SEDs show
similar peak wavelengths. For $\epsilon=0$, the wavelength ratio is 1.4
(830\AA\ vs. 1170\AA), as expected from their $T_{\rm max}$
ratio of 1.4. The $M=10^9M_\odot$ SED is nearly identical to the
solution with no wind, as expected from the negligible $\dot{M}_{\rm
  wind}$.  The factor of 10 in the $\nu_{\rm peak}$ position
(210\AA\ vs. 2130\AA) for the local blackbody solution of
$M=10^7M_\odot$ and $M=10^9M_\odot$ with a fixed $\Mdot$, is reduced
to a factor of $<3$ (830\AA\ vs. 2250\AA). At long enough wavelengths
($\lambda >3000$\AA) the SED remains unchanged, as the emission
originates from outer colder regions in the AD where the wind is
negligible.

Similar conclusions hold for the models assuming $\epsilon=1$.  In
this case the $M=10^7M_\odot$ and $M=10^8M_\odot$ models have a peak
wavelength ratio of 1.45, which is somewhat less than their $T_{\rm
  max}$ ratio of 1.65.  This difference is a result of the shallow
$T_{\rm eff}(r)$ profile in Figure \ref{f:mass_teff}. Even though
$T_{\rm eff}$ still rises as $R$ declines, the rise is small, and the
maximum occurs at smaller
radii with lower emitting area, and thus contributes relatively little to
the SED. Therefore, the ratio of the peak emission wavelengths is set
by somewhat larger radii where the $T_{\rm eff}$ ratio between the two
different mass models is smaller.  Models with $\Sigmadot(F,g)$ (not
shown) are too cold to correspond to the observed SEDs.

\begin{figure}
\includegraphics[width=84mm]{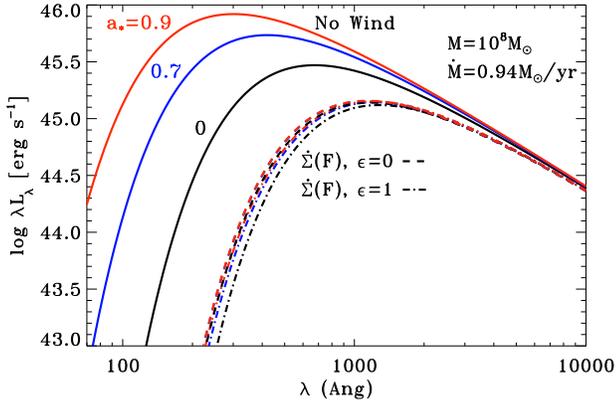}
\caption{
The specific luminosity $\lambda L_\lambda$ versus $\lambda$ for
blackbody emission corresponding to the models with different $a_*$
presented in Figure \ref{f:spin_teff}. Even though the SEDs for models without mass loss
(solid) vary significantly with $a_*$, the models with mass loss are
essentially identical. This reflects the thermostatic effect of
$\Sigmadot$, which produces a nearly isothermal disc at $r<20$ (see
Fig. \ref{f:spin_teff}).  The SED is thus blind to the value of $a_*$.}
\label{f:comp_spec_spin}
\end{figure}

Figure \ref{f:comp_spec_spin} presents the dependence of the SED on
$a_*$ based on the models shown in Figure \ref{f:spin_teff}. The
models with no mass loss show a harder SED with increasing $a_*$, as
$r_{\rm ISCO}$ gets smaller and the AD reaches a higher $T_{\rm
  max}$. In sharp contrast, the dependence of the SED on $a_*$
disappears completely once $\Sigmadot$ is included.  Although the thin
AD still extends down to $r_{\rm ISCO}$ (see Fig. \ref{f:spin_teff}),
this innermost region is not hotter than the outer regions for the
$\epsilon=0$ models, due to the thermostatic effect of $\Sigmadot$ on
$T_{\rm max}$ mentioned above (\S 5.1). The models with $\epsilon=1$
have a shallow rise in $T_{\rm eff}$ towards the center, but the small
emitting area again means these hotter regions contribute very little
to the overall emission.  In both cases the contribution to the SED
from the $r\sim 2$ region is negligible compared to the contribution
of the $r\sim 20$ region. The SED is thus blind to the inner extension
of the disc, and therefore to the value of $a_*$, for the AD
parameters explored in this figure.

\subsection{The TLUSTY models}

Due to atomic features and electron scattering it is expected that the
local SED may differ significantly from blackbody emission
\citep[SS73;][]{1984AdSpR...3..249K}.  Detailed modeling of the disc
vertical structure is required to accurately model these departures
from blackbody emission (see e.g. Hubeny et al. 2000; 2001).  When
mass loss is an appreciable fraction of the mass accretion rate, a
substantial portion of the disc surface layers can no longer be in
hydrostatic equilibrium.  In the CAK theory these departures from
hydrostatic equilibrium are due to the force multiplier from line
opacity.  In principle these lines could be modelled directly by a
stellar atmospheres code such as TLUSTY, but this would significantly
increase the complexity and computational cost of such calculations
(e.g. Kudritzki \& Puls 2000).  Therefore, we approximate the disc
emission using hydrostatic models, as used in
previous studies with no mass outflow \citep{hub00}.  Although the
characteristic peak energy of the SED should be reasonably
insensitive to this assumption, the spectrum of emission at shorter
wavelengths may be significantly modified.

Even with hydrostatic models, the mass loss has a significant impact
on the spectrum through its modification of $F(r)$ and $\Sigma(r)$, as
non-blackbody models are generally sensitive to both.  We now proceed
using the same integration method described in section 5.1, again
assuming no torque at the inner boundary.  We combine the resulting
profiles of $F(r)$ and $\Sigma(r)$ with a radial profile of the
vertical gravity, and use the interpolation methods described in
\cite{2006ApJS..164..530D} to construct full AD SEDs,
accounting for relativistic effects on photon geodesics
\citep{2009ApJ...696.1616D}.  We compute $\Sigma(r)$ using an $\alpha$
relation for the stress and solving an algebraic equation that
smoothly transitions between the gas and radiation pressure dominated
limits, as described in Appendix B of \cite{2012MNRAS.424.2504Z}.  The
only difference here is that we compute $F_1=W_{R \phi}/\alpha$ using
our numerically integrated $W_{R \phi}$ rather than assuming that the
last equality in their equation  (B3a) holds.

\begin{figure}
\includegraphics[width=84mm]{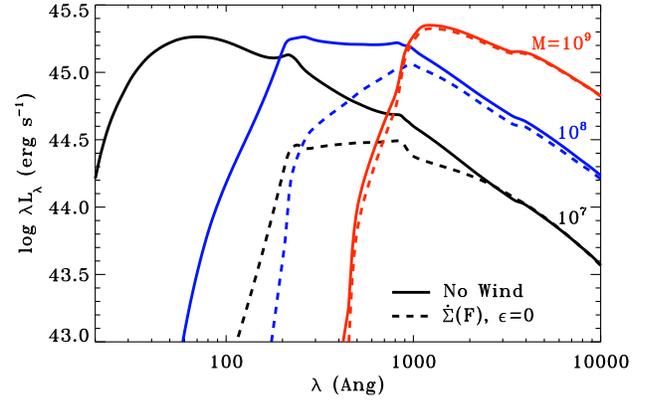}
\caption{Comparison of the specific luminosity $\lambda L_\lambda$
  versus $\lambda$ for TLUSTY-based models with (dashed) and without
  (solid) mass loss for different $M$ values.  Mass loss assumes 
$\Sigmadot(F)$ and $\epsilon=0$, with
  each model computed to yield $\Mdot(r_{\rm out})=0.94 M_\odot/{\rm yr}$.
  All models account for relativistic
  effects on photon propagation, and assume an inclination of $40^\circ$.  
The TLUSTY based models are much
  harder due to modified blackbody effects resulting from significant
  electron scattering opacity and the presence of strong edge features.
  All the models with mass loss show a peak and a spectral break
  near 1000 \AA, similar to the universal break at 1000 \AA\ seen in
the spectra of AGN. However, the spectral slope at $\lambda<1000$~\AA\ is 
predicted to be harder at lower $M$ models.}
\label{f:tlusty_seds}
\end{figure}

Figure \ref{f:tlusty_seds} compares the TLUSTY derived SEDs as a
function of $M$ for models with and without mass loss.  All models
have $\Mdot(r_{\rm out})=0.94 \rm M_\odot/yr$, $a_\ast =0$,
$i=40^\circ$.  Here we only consider one mass loss prescription, using
the $\Sigmadot(F)$ relation with $\epsilon=0$ as an example. The
TLUSTY based models with no mass loss peak at higher energies compared
to the local blackbody models of the same parameters
(Fig. \ref{f:comp_spec_mass}), due to the larger fraction of the
opacity dominated by electron scattering at short wavelengths, and the
resulting modified blackbody emission.  Absorption edge features are
also present. All the models with mass loss are significantly colder,
as expected. An interesting new feature is that, in contrast with the
local blackbody models, where there is a gradual shift to lower
$\nu_{\rm peak}$ with a rising $M$, here all models show a similar
peak position near the Lyman edge, and a break in the spectral slope
above and below the edge. This occurs because of the jump in
absorption opacity across the edge.  This spectral break is remarkably
similar to the universal break observed at $\lambda\sim 1000$\AA\ in
the mean SEDs of AGN (Telfer et al. 2002; Shull et al. 2012). The
slope shortward of 1000\AA\ depends on $M$, and the lowest
$M=10^7\Msun$ model shows that the spectral slope can remain close to
$-1$ far into the EUV.

\subsection{The derivation of $\Mdot$}

\begin{figure}
\includegraphics[width=84mm]{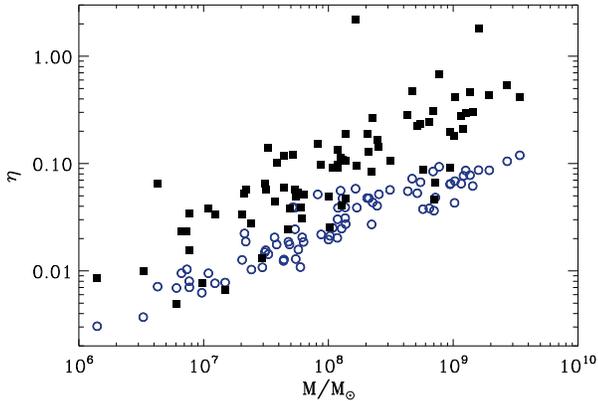}
\caption{The radiative efficiency $\eta$ versus $M$ for a sample of 80
  PG quasars.  The black filled squares denote the efficiencies inferred by
  DL11, assuming a standard disc model with no mass
  loss to estimate $\Mdot$.  The blue open circles denote the implied
  efficiencies in models with mass loss given by $\Sigmadot(F)$
  for the same $M$ and $\Mdot$ used in the DL11 analysis.  They assume $a_*=0.9$.  The
  slope of the correlation implied by the models is qualitatively
  consistent with the DL11 results, but with a lower
  mean and reduced variance of $\eta$.}
\label{f:efficiency}
\end{figure}

How does the modified AD SED derived here affect the derivation of
$\Mdot$ from the optical luminosity in AGN?  In an earlier work
(DL11), we provide a useful expression which allows to derive $\Mdot$
based on the optical luminosity, when $M$ is known, assuming the
continuum is produced by a thin AD with no mass loss. As
shown above (excluding some implausible very cold AD models produced
by the $\Sigmadot(F,g)$ relation), the wind becomes significant only
in the hotter UV emitting regions, and thus it has a negligible effect
on the optical emission. Thus, the AD based $\Mdot$ derivation remains
valid. This method was applied to the PG sample of quasars to derive
the radiative efficiency $\eta\equiv L_{\rm bol}/\Mdot c^2$, which was
found to show a clear trend with $M$ of the form $\eta \propto
M^{0.5}$ (DL11). This trend can be interpreted as an indication for a
trend of a rising $a_*$ with $M$. However, as discussed in DL11, the
$\eta \propto M^{0.5}$ relation can be derived from the universal SED
of AGN, and it was not clear whether the trend of $\eta$ with $M$ just
happens to lead by coincidence to a universal SED, or whether the
universal SED is a more fundamental property, which leads to an
apparent $\eta \propto M^{0.5}$ relation. Now that we have a physical
mechanism which may lead to a rather universal SED, we can explore its
effect on the $\eta$ versus $M$ relation, and in particular study
whether the observed $\eta$ versus $M$ relation has any implication on
an $a_*$ versus $M$ relation.

Figure \ref{f:efficiency} presents the $\eta$ vs. $M$ relation derived
in DL11 for the PG sample of quasars. We now explore whether a similar
relation can be derived if the SED of these objects is produced by an
AD with mass loss, using the $\Sigmadot(F)$ relation with
$\epsilon=0$, but with a fixed value of $a_*$. For each object we
construct a local blackbody AD+wind model with the tabulated $M$ in
DL11, and a fixed value of $a_*=0.9$ for all objects.  We iterated
over the boundary value of $\Mdot(r_{\rm in})$ in order to get the
tabulated $\Mdot$ for that object in DL11. We then integrated over the
AD luminosity to get the predicted $L_{\rm bol}$, and from that derive
the predicted radiative $\eta$, which is plotted in Figure
\ref{f:efficiency}. The $\eta$ derived here tends to be lower by a
factor of 2-3 from the one measured in DL11, however it shows a rather
tight correlation with $M$, of a similar slope to the one derived in
DL11.  Since the $\eta \propto M^{0.5}$ correlation can result from a
universal SED, it is not surprising that the disc+wind model SED used
here to derive $L_{\rm bol}$, leads to a similar relation, as this
model produces similar SEDs over a wide range of $M$ and $\Mdot$. This
trend can also be understood from the simplified analytic solution
(equation  12), which gives $r_{\rm eq}\propto M^{-0.48}$, and the fact that
in a thin AD $\eta$ is given by the innermost disc
radius. Thus, the observed $\eta$ vs. $M$ relation in DL11 does not
necessarily imply an $a_*$ vs. $M$ relation, as $\eta$ may be set
by $r_{\rm eq}$, which is independent of $a_*$.

\section{Discussion}
\label{discussion}

The SS73 solution was constructed with stellar-mass black hole systems in mind, which have a 
much hotter AD than in AGN. As a result, 
the dominant opacity in stellar systems is electron scattering and free-free absorption.
In AGN AD, the maximum temperature drops from $\sim 10^7$ to $\sim 10^5$K, and UV line opacity becomes
the dominant photospheric opacity source. This likely explains why the observed SED in
binary black hole systems is well matched by simple thin AD models
(Davis et al. 2005; 2006), while in
AGN there is generally a gross mismatch between the predicted and observed emission from the
innermost part of the AD. Various AGN AD models did take bound-free opacity into 
account (e.g. Czerny \& Elvis 1987; Laor \& Netzer 1989; Ross et al. 1992; Storzer 1993; 
Sincell \& Krolik 1997), and also included careful calculations of the vertical structure 
coupled to the radiative transfer (Hubeny et al. 2000; 2001). However, none of the models 
included line opacity. In O stars, the line opacity inevitably leads to a wind, and it may 
have a similar effect in AGN.

Winds are prevalent in AGN, as indicated by the broad and blueshifted
resonance line absorption observed in broad absorption line quasars
(e.g. Reichard et al. 2003).  The winds most likely originate from the
AD, and a likely driving mechanism is radiation pressure on resonance
lines, as indicated by both analytic solutions (Murray et al. 1995)
and numerical calculations (Proga et al. 2000). Here we find that
radiation pressure driven winds may modify significantly the disc
structure. Applying the mass loss per unit area measured in O stars,
we find that the simple thin disc solution effectively terminates at a
few tens of $R_g$.  The steep dependence of the local mass loss
on $T_{\rm eff}$, $\Sigmadot(F)\propto T_{\rm eff}^{7.6}$ (equation 2)
sets a cap on the maximum $T_{\rm eff}$, which is well below $10^5$K. 
This can explain why the observed AGN SEDs do not show a rise
towards the EUV, and may also explain the rather universal turnover
observed at $\lambda<1000$\AA.


The softening effect of an AD wind mass loss on the SED of AGN was
noted by Witt et al. (1997) and Slone \& Netzer (2012), and
for cataclysmic variables by Knigge (1999). The new result here is the
application of the stellar mass loss to AGN AD, which yields a UV
turnover similar to the one observed, with no free parameters. In
contrast with Slone \& Netzer (2012) and Knigge (1999), where
general $\dot{M}_{\rm wind}(r)$ relations were assumed, here we find that
the derived $\dot{M}_{\rm wind}(r)$ relation which matched the
observed SED, has a negligible effect on the derived $\dot{M}(r_{\rm
  out})$ (DL11), as there is negligible wind at the region where the
optical emission is produced. However, it is certainly correct that
$\dot{M}(r_{\rm in})$ may be significantly smaller than
$\dot{M}(r_{\rm out})$ , as was pointed out by DL11 and Slone \&
Netzer (2012).

\citet{2012MNRAS.423..451L} proposed that obscuration near, but
external to the accretion disc, produces the nearly constant SED peak.
Obscuring material off the plane of the disc but at low radius is
assumed to be provided by outflows from (or instabilities in) the
disc.  This model assumes that mass lost to obscuring clouds is not
large enough to modify the intrinsic disc emission, which is
instead altered by transfer through the obscuring clouds.  Although we have not computed
the effects of reprocessing by the outflow, such reprocessing likely
occurs and the effects described in \citet{2012MNRAS.423..451L} may be
present at some level.

\subsection{Are O stars winds applicable to AGN?}
 The CAK wind
solution applies for spherically symmetric systems, where both gravity
and the radiation field fall off as $1/r^2$.  In AD there are
significant differences. Thin AD are rotationally supported, and in
the local disc frame, close to the disc surface, gravity increases
linearly with height, while the radiation field is independent of
height. Thus, it is not clear that the CAK solution is relevant even
locally close to the surface of the AD. Far enough above the disc, where
$h>r$, both
gravity and the radiation field become radial and fall off as $1/r^2$. 
Thus, in contrast
with the stellar case, in AD both the relative strength and 
directions of the radiative and gravitational forces change with
position.  Proga, Stone \& Kallman (2000) find that as a result there
is no steady state solution, in contrast with the steady state stellar
wind solution.

A related issue is the difference in the velocity fields in AD and in
O stars.  AD are expected to have Keplerian velocity shear and be
highly turbulent. Since line driven wind models from O stars generally
assume monotonically increasing velocity profiles, non-monotonic
velocity distributions (due e.g. to the turbulence) may modify the
acceleration of the outflow.  Modeling such effects requires fairly
sophisticated radiative transfer calculations beyond the scope of this
work, which to the best of our knowledge have not been considered
elsewhere.

In addition, the integrated SED in AGN is harder than in stars, and
thus the wind is subject to over-ionization once it becomes exposed to the
harder EUV - soft X-ray radiation, which may shut off line driving if
the ionization is large enough.  Previous
studies (e.g. Murray et al. 1995) have generally assumed that such
irradiation only becomes important after the matter has been lifted
significantly above the disc surface.  However, if the inner region of
the disc is thick enough it may directly irradiate the disc at
larger radii, increase the ionization level and 
reduce the force multiplier at the disc surface. Such direct
irradiation is not expected in the radiation dominated
regions of a SS73 solution because the AD scale height is nearly constant
with radius.  In any case, the irradiating flux falls off a $r^{-3}$ and
likely remains a small fraction of the locally dissipated flux (Blaes 2004).  
Significant flaring of the disc or
scattering of radiation by the failed wind might increase the
irradiation, enhancing the ionization level at the disc surface
and potentially inhibiting the outflow.  For the line driven wind to be a
viable explanation of the `universal' UV break, the outflow clearly
cannot be completely quenched. 

The local force multiplier within
an unirradiated disc atmosphere is likely similar to that in an
atmosphere of a star with the same $T_{\rm eff}$. The force multiplier
is $\Gamma \sim 10^3$ (Lamers \& Cassinelli 1999) at a column of $\sim
10^{18} \; \rm cm^{-1}$ from the disc surface, which inevitably leads
to a modification of the disc vertical structure. The SS73
solution assumes only electron scattering, leading to a thin disc with
a height $h<r$ for $\mdot<1$.  An increase in the opacity, i.e. $\Gamma>1$, will
expand the disc atmosphere vertically, and can lead to 
a disc thickness $h'\sim \Gamma h$. For $\Gamma>r/h$ the disc atmosphere becomes
geometrically thick. If $h'>r$ then $g\propto 1/r^2$, a hydrostatic solution is not
possible anymore, and a wind is launched. Since $r/h\sim
10-100$ typically in thin discs, a $\Gamma \sim 10^3$ should lead
to a wind. However, once the wind is exposed to the central ionizing
continuum, it may get significantly ionized (depending on its density), 
leading to $\Gamma \la 10$ (Murray et al.  1995, Fig.9
there). An acceleration length of $\sim r/\Gamma$ will bring the wind to the
escape velocity, and allow it to escape, even if it gets
over ionized at the coasting phase.
If the outflow does not attain the escape speed before being over ionized, 
the gas will fall back to the disc, forming a `failed wind'.


Proga et al. (2000) present a model for a UV line driven wind from an
AD with $m_8=1, \mdot=0.5$. They find a time averaged $\Mdot_{\rm
  wind}/\Mdot=28$\%, for $\Mdot_{\rm wind}$ measured at $r>200$.
This wind is shielded by the failed wind produced close to the inner
boundary of the simulation (at $r=150$), which is over-ionized by the
central X-ray source. The AGN AD wind simulation of Proga \& Kallman
(2002) show a sharp rise in the local mass loss at $r<100$ (fig.1
there), and thus the integrated $\Mdot_{\rm wind}$ lifted from the
surface may reach $\Mdot_{\rm wind}/\Mdot\sim 1$ closer to the center,
as derived by our simplistic use of the CAK solution. As noted by
Proga (2005) this line driving can change the vertical structure of
the disc, and likely produce a ``puffed up'' disc.

\subsection{What happens in the innermost disc?}

In section \ref{results}, we considered two sets of models which can
have rather different implications for mass loss from the disc.
Models in which the work done in unbinding the outflow is negligible
($\epsilon=0$) can nominally drive almost all of the matter out of the
{\it thin disc}.  However, the energy required to drive all
this material to escape velocity can exceed the energy released 
by the remaining disc material which accretes to the center 
(e.g. the $10^7 \Msun$ models with $\dot{m}_{\rm
  out}= 2$, see Fig.3).  Hence, there must be a substantial failed wind in this
scenario and most of the material in this failed wind must eventually
accrete, although possibly not in the form of a standard, thin disc.

In the second set of models with $\epsilon=1$, we account for the
energy lost by radiation in unbinding the outflow to compute $F$ and
use this to evaluate the $\Sigmadot$ relations.  In this case the mass
loss is limited by the corresponding reduction in $F$. The mass
outflow rate can still be quite large ($\sim 90$ \% of $\dot{M}$ at
large radius), but the accreted mass is significantly larger relative to the
$\epsilon=0$ models.  All the energy needed to accelerate the outflow
to escape velocity is accounted for and there is no need for a failed
wind to form a hotter, radiatively inefficient flow.

Although the latter models have the appeal of being explicitly energy
conserving, the numerical simulations discussed above (e.g. Proga et
al. 2000) suggest winds may not operate in this fashion.  They are
predominately accelerated by the radiation from annuli interior to
their launching radius, so the assumption of a constant $\epsilon$ is
likely a poor approximation.  More importantly, the winds behave (in
certain respects) more like the $\epsilon=0$ models, in that they
launch more matter from the thin disc than they accelerate to infinity
and forming a ``failed wind.''  These simulations assume a constant
$\Mdot$ AD as a boundary condition, and find outflow rates
as high as 50\% of this prescribed $\Mdot$.  In principle, even higher
outflow rates could be obtained, but such models have not been
considered due to the assumed lack of feedback on the boundary condition
in the models used (D. Proga, private communication). 

Our best guess is that a real system would behave like
some combination of the above models: a significant fraction of the
radiative flux will go into accelerating some fraction of the outflow
that does exceed escape velocity and becomes a wind, but not all of the
mass removed from the thin disc will become unbound.  Much of it may
form a failed wind that returns to the disc or accretes through a
hotter, geometrically thick flow.

The distinction between the thin disc and the outflow probably breaks
down when a non-negligible fraction of the disc is no longer in
hydrostatic equilibrium. Since the calculations presented in section 5
do not account for the detailed vertical structure, the radial
distributions of $T_{\rm eff}$ and $\Mdot$ in regions where
$\Mdot_{\rm wind}(r) \sim \Mdot(r_{\rm out})$ should be interpreted with
this in mind.  If most of the material lifted from the disc falls
back, say due to over ionization, then it must eventually all accrete
to the center. In this case, observations would seem to require a hot,
radiatively inefficient flow, since substantial thermal EUV emission
is not seen.  The bulk of ``fallback'' flow must remain hotter and
thicker than standard solution, cooling primarily via inverse-Compton
scattering and contributing significant radiation only in the X-ray
band.  In this scenario, line-driving would result in a transition
from a thin, radiatively efficient disc to a geometrically thick,
radiatively inefficient flow.

A low radiative efficiency can occur in hot flows if the optical depth
is very low or very high
\citep[see e.g.][]{1994ApJ...428L..13N,1995ApJ...444..231N,
1988ApJ...332..646A,1995ApJ...438L..37A}.
When the optical depth is large, the radiation is advected inwards
inside a thick disc, when the radiation diffusion time-scale is
longer than the infall time-scale. This effect is expected to be
present in discs with $\mdot>1$, where the AD becomes slim, rather
than thin \citep{1988ApJ...332..646A}. Interestingly, all objects in
DL11 with an observed radiative efficiency below 2\% have measured
$\Mdot$ which corresponds to an AD with $\mdot\sim 3-30$
for a 10\% efficiency. So, advection of radiation may indeed suppress
the emission from the innermost disc. However, the SED of such a high $\mdot$ 
disc typically peaks well shortward 1000\AA, so it is inconsistent with the
observed $\lambda\sim 1000$\AA\ turnover. Another process is responsible
for the universal $\lambda\sim 1000$\AA\ turnover.

The unbound material in the inner AD must remain hot enough to avoid
emitting significantly in the observable UV band.  What is the
expected emission from this hot inner region?  Can it form the
thick and hot inner structure often invoked as a source for the X-ray
emission?  Let us assume the X-ray emission is produced in the inner
AD region through comptonization 
of the incident thin disc emission which comes from $r>r_{\rm eq}$, where
the thin disc resides. 
The Compton cooling takes place only in
a $\tau_{\rm es}\sim 1$ surface layer of the illuminated hot and thick
configuration at $r<r_{\rm eq}$. So, only a small fraction of the inner 
disc volume will cool radiatively, while the rest of the dissipated energy 
should be
advected inwards. The estimated X-ray luminosity of this surface layer
can be derived based on the thermal energy stored in this layer divided
by the cooling time. 
The likely projected surface area of the inner hot disc 
is $A=cos(\theta) 2\pi r^2$, where $r\sim r_{\rm
  eq}$ (equation 12), and $cos(\theta)\sim 0.1$ is the illumination angle
of the external radiation, which yields $A=1.6\times 10^{29}
m_8^{1.04}\Mdott^{0.48}$~cm$^2$. The illuminated column density corresponds
to $\tau_{\rm es}\sim 1$. The electron temperature is $T\sim 10^9$K (from
the spectral slope of $-1$ for comptonization, Rybicki \& Lightman
1979, equation 7.45b). Thus, the total thermal energy of the electrons in
the illuminated layer is $E=A\Sigma kT=5\times
10^{46}m_8^{1.04}\Mdott^{0.48}$~erg.  The Compton cooling time of
electrons embedded in a blackbody at $T=10^5T_5$ is $t_C=81.5
T_5^{-4}$~s (e.g. Laor \& Behar 2008, equation  53).  Approximating the disc
emission as a blackbody at
$T_{\rm max}$ (equation 14), gives a cooling rate of the illuminated layer of
$E/t_C=L_X\sim
10^{44}\Mdott^{0.76}m_8^{0.48}$.  How does it compare with the
bolometric luminosity? The expected $L_{\rm bol}$ from the
geometrically thin AD, which extends down to $r_{\rm eq}$, is $\sim
\Mdot c^2/r_{\rm eq}$, or $L_{\rm bol}\sim
10^{45}\Mdott^{0.76}m_8^{0.48}$. We thus get $L_X\sim 0.1 L_{\rm
  bol}$, which is close to the typical ratio observed in AGN.

To summarize, the radius of the inner thick disc, the assumption it is
maintained at $T\sim 10^9$K by the viscous dissipation, the implied 
Compton cooling time based on
the incident radiation from the thin AD, and the thermal energy
content of the cooling surface layer of the inner hot disc, happen to
combine together and give a constant $L_X/L_{\rm bol}$, of
the order of magnitude observed. 

If the inner X-ray source is produced by a breakdown of the thin disc
solution due to line opacity driving, while the X-ray source is
significant enough to quench the disc wind, this may lead to an
instability in the X-ray and EUV emission, and an
anticorrelation between the two bands.  The inner AD may switch
back and forth from a thermal EUV emitting thin AD state, which
develops a strong wind and turns into a thick and hot X-ray emitting
configuration, which shuts off the wind, and turns back to the thin
thermal EUV emission state, which again develops a strong wind.

\subsection{What happens at lower Masses?}

With decreasing $M$ the disc
gets hotter, the wind starts at a larger $r$ (eqs.12, 21), and the
radiative efficiency is expected to get lower (see also 
$\eta \propto M^{0.5}$, DL11). This may partly explain why 
accreting $M \le 10^6\Msun$ systems are rare. A wind launched at a larger $r$ 
may find it easier to escape, and low $M$ black
holes may find it harder to grow by accretion.  However, at a low
enough $M$, $r_{\rm eq}$ will move out into the gas pressure dominated
regime of the disc. In this regime the disc is vertically supported by
the gas pressure, so an increase in radiation pressure due to
$\Gamma>1$ does not necessarily lead to a significant change in the vertical
structure. A values of $\Gamma>10^3$ close to the disc surface will
most likely overcome gravity and drive a wind. However, in contrast
with the uniform vertical density profile of the radiation dominated
part of the disc, in the gas dominated part the density drops steeply with height,
and thus the density at the sonic point, which feeds
the base of the wind may be significantly lower, lowering $\Mdot_{\rm wind}$. 

Another effect which comes in with decreasing $M$ is that $T_{\rm max}$ increases,
reaching $>2\times 10^5$~K for $M=10\Msun$ (eqs. 15, 23). At this temperature
$\Gamma$ likely decreases due to over-ionization of the primary atomic 
UV absorbers, which will also reduce $\Mdot_{\rm wind}$.  

Observations of Cataclysmic Variables (CV) which harbour AD around white dwarfs, 
yield winds with $\Mdot_{\rm wind}<<\Mdot$ (e.g. Feldmeier et al. 1999), 
despite the fact that the CV AD peaks in the UV. However, these systems are
characterized by $\Mdott\sim 10^{-8}$, $m_8\sim 10^{-8}$, and 
 $r_{\rm in}\sim 10^4$. Plugging these values into equation (11) gives 
$\Mdot_{\rm wind}/\Mdot\sim 10^{-2}$, which is likely an overestimate as the
AD in CV is gas pressure rather than radiation pressure dominated. Thus, 
CV AD have only weak winds, despite their peak UV emission and the associated high
$\Gamma$ values, due to their large $r_{\rm in}$ (see equation 11).

In X-ray binaries (XRB) systems, the disc is much more compact, reaching $T\sim 10^7$~K.
Therefore, the gas becomes fully ionized and line opacity is negligible. If
the outer disc extends far out enough, it can reach the $\Gamma\gg 1$
regime, this time from the other side of the opacity barrier, probably
at $T< 10^6$~K, where the line opacity starts to build up, which
may also produce a wind, this time from the outer disc, rather than
the inner disc.  The wind is expected to move inwards in the low hard
state, as the disc gets cooler, which may disrupt the thin disc formation
down to the center.

For an $M<10^6 \Msun$ black hole
radiating near Eddington, the inner regions of a SS73 disc reach
$T> 10^6$~K, and should emit
predominantly in the soft X-rays \citep{2012MNRAS.420.1848D}.  
At this temperature the ionization may be high enough to reduce
$\Lambda$ and allow a thin disc solution with no wind. Further out the disc will
be colder, and significant mass loss is likely to occur, which may 
strongly suppress $\Mdot$ which can arrive to the center. However, if a 
fraction of $L_{\rm bol}$, which peaks at soft X-rays, 
is intercepted farther out in the disc,
wind launching may be quenched further outside.  Such a
scenario may explain some low mass ($M <10^6 \Msun$) narrow line
Seyfert 1s with large soft X-ray ``excesses'', such as 
RE J1034+396 \citep{2012MNRAS.420.1848D}. 

\subsection{Can $a_*$ be measured from the SED and the Fe K$\alpha$ line?}

Since the line driven wind effectively terminates the thin disc well outside
$r_{\rm ISCO}$, the value of $a_*$ does not affect the thin disc SED. Thus, in contrast with
XRB, we generally do not expect the AGN SED to provide a useful constraint on $a_*$.
However, for a sufficiently cold disc, with a turnover at $\lambda>1000$\AA,
the wind disappears. This may explain the remarkably good fit of a simple local BB AD model 
to SDSS J094533.99+100950.1,
(Czerny et al. 2011; Laor \& Davis 2011), which is a weak line quasar with a UV turnover at
$\lambda\sim 2000$\AA\ (Hryniewicz et al. 2010). We therefore expect that other cold AD quasars,
i.e. quasars with a blue SED at $\lambda>3000$\AA\ (which excludes dust reddening), and a UV turnover
at $\lambda>1000$\AA, can also be well fit by a simple local BB AD model. One may therefore be able
to constrain $a_*$ based on the SED in these quasars, as commonly done now in XRB
(e.g. Li et al. 2005; Shafee et al. 2006; McClintock et al. 2011)).


Similar issues may apply to efforts to measure $a_*$ via models of the
Fe K$\alpha$ line and ionized reflection.  The numerical solutions
indicate that an optically thick disc extends down to
$r_{\rm ISCO}$. However unless the disc is cold, only a small fraction
of $\Mdot$ extends down to $r_{\rm ISCO}$. The numerical solution
ignores the material which left the thin disc, which may form a thick
configuration in which the thin disc is embedded, if it exists at
all. A lack of a thin cold bare disc which extends down to $r_{\rm
  ISCO}$ will affect the expected profile of the fluorescence Fe
K$\alpha$ line, produced by X-ray reflection from the AD.  The line
may be dominated by emission outside $r_{\rm eq}$, which will make the
line narrower, and its profile independent of the value of $a_*$. Line
emission inside $r_{\rm eq}$ will depend on the gas temperature, and
the line profile may be modified by possible scattering effects at a thick
and hot surface layer. In any case, the emission is not expected to originate
from a thin bare AD, as commonly assumed, as a thin bare disc which
extends down to $r_{\rm ISCO}$ is simply not observed. We do predict
that the colder the disc is, the further it extends inwards, and the
broader the K$\alpha$ line can be.

\subsection{What is the effect on the characteristic half-light radius?} 
The outflow changes the radial distribution of the emitted flux in
different frequency bands.  In particular, the reduction of the far UV
emission relative to the SS73 solution leads to an increase in the
half-light radius for the optical to UV bands.  This is because the
Rayleigh-Jeans tail of the emission from the UV peaked regions of the
disc contributes a non-negligible fraction to the overall SED at
longer wavelengths.  Removing this emission or shifting it to X-ray
wavelengths increases the fraction of the long wavelength emission
from larger radii.

This effect is interesting in light of claims based on microlensing
analysis of lensed quasars that the optical to UV emission comes from
radii which are factors of $\sim 3-10$ larger than expected from the
standard thin disc model
\citep[e.g.][]{2005ApJ...628..594M,2007ApJ...661...19P,2010ApJ...712.1129M}.
However, the increases in the half-light radius (which the
microlensing results are claimed to measure) that we infer are
typically $< 30\%$ at 2000 \AA~and $< 10\%$ at 4000 \AA~relative to
the models with no outflow, so this effect cannot account for the
large discrepancies that are claimed.  A caveat is that we have not
considered the reprocessing of the thin disc emission by the outflow.
For such large mass loss rates the wind will likely
be optically thick to Thomson scattering in the radial direction
\citep{2010MNRAS.408.1396S}.  Some fraction of this will be scattered
downward and be reprocessed by the disc at larger radii.  This
reprocessed emission should increase the half-light radius at longer
wavelengths, although a more detailed calculation is required to
estimate the possible magnitude of the effect.

\subsection{Predictions}

The inwards extent of the AD can be probed through the position of the
UV turnover.  The $\Sigmadot(F)$ wind relation generally predicts a
turnover at $\lambda_{\rm peak}\sim 1000$\AA. However there is still
some residual dependence on $\mdot$ and $m_8$, possibly in the form of
$T_{\rm max}\propto (\mdot/m_8)^{0.07}$ (equation  15), which should be
observationally detectable. The quantity $\mdot/m_8$ can be estimated
based on the broad emission lines, which gives $M=v_{\rm BLR}^2 R_{\rm
  BLR}/G$, where the H$\beta$ FWHM is used to derive $v_{\rm BLR}$. In
addition, $R_{\rm BLR}\propto L^{0.5}$ based on reverberation
mappings, and the above relations imply $\mdot/m_8\propto v_{\rm BLR}^4$.  Thus, the
$\Sigmadot(F)$ wind relation (equation 15) implies that $T_{\rm max}\propto v_{\rm
  BLR}^{0.28}$, or $\lambda_{\rm peak}\propto v_{\rm BLR}^{-0.28}$. A
factor of 10 increase in $v_{\rm BLR}$ will therefore be associated
with a decrease by factor of 2 in $\lambda_{\rm peak}$. This is true
as long as the disc is not too cold to become windless, i.e. when
$v_{\rm BLR}$ is not too large ($\la 10,000$~km~s$^{-1}$, e.g. Laor \&
Davis 2011).  In general, the detection of a correlation of $\lambda_{\rm peak}$
with $\mdot$ and $m_8$ can provide a hint on the specific form of $\Sigmadot$
in AGN AD.

A clear prediction is a metallicity dependence of $\lambda_{\rm peak}$. 
Radiation driven wind models for hot stars
predict a close to linear relation of $Z$ and $\Mdot$ (e.g. Abbott 1982; Leitherer et al. 1992; 
Vink et al. 2001; Kudritzki 2002). Since $r_{\rm eq}\propto \Mdot_{\rm wind}^{0.27}$  for the $\Sigmadot(F)$ 
wind relation, and in AD $T_{\rm max}\propto r^{-3/4}$, we expect that
$\lambda_{\rm peak}\propto Z^{-0.2}$. Although the expected 
dependence of $\lambda_{\rm peak}$ on $Z$ is weak, it is a robust prediction
of line driven winds, and its detection will provide strong support for the wind interpretation
put forward in this paper.

\section{Conclusions}

We derive the Navier-Stokes equations for gas on circular orbits in a thin AD,
including local mass loss and angular momentum loss terms. We then solve these equations
numerically for a Keplerian disc, with zero torque inner boundary condition.
We apply the local mass loss terms per unit area measured in hot stars, as a function 
of $F$ and $g$, at the AD surface. We assume no angular momentum loss induced by the
mass loss. We calculate the derived SED based on the local blackbody approximation,
and also using the TLUSTY stellar atmosphere code. We find the following.

\begin{enumerate}

\item Line driven winds put a cap on the AD $T_{\rm max}<10^5K$, with a 
weak dependence on $M$ and $\mdot$. This cap is
consistent with observation of AGN SEDs, which generally show a spectral turnover near
1000\AA. 

\item In most cases the thin AD is effectively truncated at a few tens $R_g$, 
well outside $r_{\rm ISCO}$.
The derived SED is thus independent of the value of $r_{\rm ISCO}$, and is therefore
independent of the value of $a_*$. 

\item In a cold AD, defined by an SED with a turnover longward of 1000\AA, the line 
driven wind is negligible. It may therefore be possible to use the SED of objects 
predicted to have a cold AD, based on their measured $M$ and $\mdot$, to derive 
the black hole $a_*$.

\item The TLUSTY based models with winds tend to show a spectral break
at $\lambda \sim 1000$\AA, due to a combination of the Lyman edge and the truncation of
the hot inner part of the AD due to the wind. 

\item Depending on the mass loss prescription, $M$, and $\dot{m}$ of
  the model, the material removed from the thin disc either escapes as
  a wind or forms a failed wind that must accrete. In either case, 
  the thin disc solution of SS73 cannot generally
  extend to $r_{\rm ISCO}$, as the emission from the inner
  region of a thin disc is generally not observed.

\item For a sufficiently large failed wind, the inner disc must be
  radiatively inefficient.  It may form a geometrically thick hot
  inflow. If the electron temperature can be maintained at $T\sim
  10^9$~K, then Compton cooling leads to $L_X\sim 0.1 L_{\rm bol}$.

\item The $\Sigmadot(F)$ wind relation implies a radiative efficiency
which scales as $M^{0.5}$, which agrees with
the measured relation (DL11). Thus, the low radiative efficiency of low $M$ AGN 
does not imply a low $a_*$, but may be induced by high mass loss and the 
implied large truncation radius of the inner thin AD.

\item If the UV turnover is indeed a line driven mass loss effect, then the effect 
is necessarily $Z$ dependent. Higher $Z$ object should show a larger 
$\lambda_{\rm peak}$, i.e. a softer ionizing SED at a given $M$ and $\Mdot$.

\end{enumerate}

Rather detailed wind simulations are available, and it will be interesting 
to explore the radiation transfer through the wind/failed wind, and its possible 
impact on the observed SED. However, a major uncertainty in wind models
remains concerning the vertical structure of the disc, in particular the top
layer from which the wind is launched.
The structure depends on the exact nature of the turbulence and how it is
dissipated. Thus, one cannot yet derive estimates for
$\Mdot_{\rm wind}(r)$ from first principles.  
We went around this major uncertainty by
adopting $\Mdot_{\rm wind}(r)$ derived based on $\Mdot_{\rm wind}(g,
T_{\rm eff})$ of O stars.  Although the models adopted here are rather
simplified, they yield SEDs remarkably similar to those generally
observed, without significant sensitivity to the details of the
launching mechanism.  However, the structure of the innermost disc and
its possible feedback, in particular when the implied outflow becomes
comparable to the accretion rate, need to be further explored.

One possible feedback is X-ray irradiation of the innermost disc,
which if strong enough, may ionize the surface disc layer to a level
which will quench the line driven wind. If the hot inner disc is
indeed formed by a line driven failed wind, then quenching the wind
may quench the X-ray source, which will allow the wind to form
again. 

Despite the associated uncertainty in the above analysis, 
a plausible statement one
can make is that in AGN, as in O stars, {\em ``a static atmosphere is
  not possible''} (CAK, \S II there). The true structure of the inner
AD remains to be understood.

\section*{Acknowledgments}
We thank the referee, Ramesh Narayan, for a critical and constructive
review, which
significantly improved the paper.  We
thank Oren Slone for pointing out that applying $\Mdot(r)$ to the SS73
solution is invalid.  We thank Norm Murray, Daniel Proga, and Ramesh Narayan for
helpful discussions.  This research was supported by the ISRAEL SCIENCE
FOUNDATION (grant No. 1561/13). SWD is grateful for financial support from the
Beatrice D. Tremaine fellowship.

{}

\appendix

\section{Relativistic Accretion Discs with Outflow}

In section 4.1, we derive the equations of conservation of mass,
angular momentum, and energy in the Newtonian limit.  Here describe
the fully relativistic generalizations to these equations The
derivation of the relativistic equations for the steady state,
axisymmetric disc without mass loss has been discussed extensively in
previous work \citep[see
  e.g.][]{1973blho.conf..343N,1974ApJ...191..499P,1995ApJ...450..508R}.
Here we generalize these derivations to include the effect of an
outflow.  These works differ in the formulation of the eqs. of hydrostatic
equilibrium, but all agree on the appropriate thin disc limit of the
equations for conservation of mass, angular momentum and energy.
These correspond to eqs. (4), (9), and (16) in \citet{1995ApJ...450..508R}.
Specifically, they are conservation of mass
\begin{equation}
(\rho U^\alpha)_{;\alpha}=0,\label{eq:massrel}
\end{equation}
conservation of energy
\begin{equation}
S^{\alpha \beta} U_{\alpha; \beta}+q^{\beta}_{~;\beta}=0,\label{eq:energyrel}
\end{equation}
and conservation of angular momentum
\begin{equation}
\rho U^\beta U^\phi_{~;\beta}+S^{\phi \beta}_{~~;\beta}+U^\phi U_\alpha S^{\alpha \beta}_{~~;\beta}=0.\label{eq:angmrel}
\end{equation}
Here $U^\alpha$ is the four-velocity, $\rho$ is the rest mass density,
$S^{\alpha \beta}$ is the stress tensor and $q^\alpha$ is radiative energy flux.
 
Now we integrate the steady state equations over $z$, initially neglecting
any energy loss associated with unbinding of the outflow.  As above, $F_M$ represent the flux
of mass out of the thin disc.  Conservation of mass is
identical to the Newtonian version (equation  \ref{eq:mass}).  Conservation
of angular momentum and energy become
\begin{eqnarray}
\frac{\partial}{\partial r}\left( \dot{M} L^\dagger -2 \pi r W_r^{~\phi}\right)
=  4\pi r L^\dagger (Q+F_M)\label{eq:angmcons} \\
\frac{\partial}{\partial r}\left( \dot{M} E^\dagger -2 \pi r W_r^{~\phi}\Omega\right)
=  4\pi r E^\dagger (Q+F_M).\label{eq:energycons}
\end{eqnarray}
We have adopted the notation of \citet{1974ApJ...191..499P}, where
$L^\dagger$ and $E^\dagger$ denote the energy and specific angular
momentum at infinity, $W_r^{~\phi}$ is the vertical integral of the
$r$--$\phi$ component of the viscous stress tensor, and $\Omega$ is
the rotation rate of a circular orbit.  Here, $Q$ represents the
(``viscous'') dissipation associated with stresses in the accretion
flow.  (In a model with no outflow, it corresponds to the flux emitted
from the disc surface.)  In analogy with equation (\ref{eq:energy}) we
define the radiative flux observed at infinity as
\begin{equation}
F = E^\dagger Q - \epsilon F_M (1-E^\dagger).\label{eq:energyinf}
\end{equation}
where
\begin{equation}
E^\dagger=\frac{1-2/r+a_*/r^{3/2}}{\left(1-3/r+2a/r^{3/2} \right)^{1/2}}
\end{equation}
Using this expression and defining $dL/dr=4 \pi r F$,
equation  (\ref{eq:energycons}) can be integrated from $r_{\rm in}$ to
infinity to find
\begin{eqnarray}
L = \dot{M}_{\rm in} (1-E^\dagger_{\rm in})
-(\epsilon -1) \int^{\infty}_{r_{\rm in}} \frac{\partial \dot{M}}{\partial r}(1-E^\dagger) dr.
\end{eqnarray}
Here $\dot{M}_{\rm in}$ and $E^\dagger_{\rm in}$ are evaluated at $r_{\rm in}$, where
we have assumed that the internal torque vanishes.  Hence the total luminosity radiated
by the disc (as observed at infinity) is just the standard disc efficiency using the mass 
accretion rate at the inner edge minus the work done accelerating the flow beyond
its escape velocity (if $\epsilon > 1$).

For our numerical solution, it is useful to rewrite the equations in terms of
 the co-moving frame quantities $W_{r \phi}$ and the Keplerian rotation $\Omega_K$.
Equation (\ref{eq:angmcons}) becomes
\begin{equation}
\frac{\partial W_{R \phi}}{\partial R}+\frac{2(R-R_g)}{R^2 A_{\rm RH}} W_{R \phi}=
\frac{\Mdot \Omega_K}{4\pi R}\frac{E_{\rm RH}}{A_{\rm RH}B_{\rm RH}}\label{eq:angmf},
\end{equation}
and equation (\ref{eq:energycons}) becomes
\begin{equation}
Q=-\frac{3}{4} W_{R\phi} \Omega_K\frac{A_{\rm RH}}{B_{\rm RH}}\label{eq:energyf}.
\end{equation}
$A_{\rm RH}$, $B_{\rm RH}$, and $E_{\rm RH}$ are functions of $r$ and $a_*$ defined
in \citet{1995ApJ...450..508R} that approach unity for large $r$.  These are
identical to eqs. (14) and (17) of \citet{1995ApJ...450..508R}, and in the limit
that $R \gg R_g$, they reduce to eqs. (\ref{eq:energy}) and (\ref{eq:angmss}).

\label{lastpage}

\end{document}